%% file: 1469.tex
\documentclass{aa}
\newcommand{\be}{\begin{equation}}
\newcommand{\ee}{\end{equation}}
\newcommand{\lb}[1]{\label{#1}}

\newcommand{\ssty}{\scriptscriptstyle}

\newcommand{\dl}{d_{\ssty L}}
\newcommand{\da}{d_{\ssty A}}
\newcommand{\dg}{d_{\ssty G}}

\newcommand{\dsa}{d\sigma_{\ssty A}}
\newcommand{\dsg}{d\sigma_{\ssty G}}
\newcommand{\doa}{d\Omega_{\ssty A}}
\newcommand{\dog}{d\Omega_{\ssty G}}
\newcommand{\ella}{\ell_{\ssty A}}

\usepackage{graphicx}
\begin{document}
\title{Cosmological distances and fractal statistics of 
       galaxy distribution}
\author{Marcelo B. Ribeiro}
\institute{Physics Institute, University of Brazil--UFRJ, CxP 68532,
           CEP 21945-970, Rio de Janeiro, Brazil; E-mail: mbr@if.ufrj.br}
\date{Received 15 June 2004 / Accepted August 2004 }
\abstract{This paper studies the effect of the distance choice in
          { radial (non-average) statistical tools} used for fractal
	  characterization of galaxy distribution. After
	  reviewing the basics of measuring distances of cosmological
	  sources, various distance definitions are used to calculate
	  the { differential density $\gamma$ and the integral
	  differential density $\gamma^\ast$ } of the dust distribution
	  in the Einstein-de Sitter cosmology. The main results are
	  as follows: (1) the choice of distance plays a crucial
	  role in determining the scale where relativistic corrections
	  must be taken into account, as both $\gamma$ and $\gamma^\ast$
	  are strongly affected by such a choice; (2) inappropriate
	  distance choices may lead to failure to find evidence
	  of a galaxy fractal structure when one calculates those
	  quantities, even if such a structure does occur in the
	  galaxy distribution; (3) the comoving distance and the
	  distance given by Mattig's formula are unsuitable to probe
	  for a possible fractal pattern as they render $\gamma$ and
	  $\gamma^\ast$ constant for all redshifts; (4) a possible
	  galaxy fractal system at scales larger than 100Mpc ($z
	  \approx 0.03$) may only be found if those statistics are
	  calculated with the luminosity or redshift distances, as
	  they are the ones where $\gamma$ and $\gamma^\ast$ decrease
	  at higher redshifts; (5) C\'el\'erier and Thieberger's (2001)
	  critique of Ribeiro's (1995) earlier study are rendered
	  impaired as their objections were based on misconceptions
	  regarding relativistic distance definitions.
          }
\maketitle
\titlerunning{Cosmological Distances and Galaxy Fractal Statistics}

\section{Introduction}

{From} the very beginnings of modern cosmology the question exists
of whether or not there will be a distance scale where the matter
distribution in the Universe becomes uniformly distributed, or if it
will always remain inhomogeneous, following, in this case, a
hierarchical clumping pattern of matter. With the adoption of the
{\it Cosmological Principle} by the majority of researchers working in 
cosmology, it became standard to equate the
former hypothesis to a statement of truth, meaning that most researchers
adopted the view that there {\it must} be a scale beyond which a homogeneous 
distribution will be reached. This is the viewpoint presented in most, if
not all, textbooks in cosmology. As a consequence the latter hypothesis,
that is, the alternative hierarchical, or inhomogeneous, viewpoint was
pushed to the sidelines. Nevertheless, and despite the historical fact
that since the beginning of modern cosmology, in the 1920s and 1930s, the
view based on the Cosmological Principle became dominant, the hierarchical
proposal managed to survive on the fringes of cosmological research for
almost a century (de Vaucouleurs 1970; Mandelbrot 1977, 1983;
Pietronero 1987; Ribeiro 1994). 
 
The modern version of a hierarchical distribution of matter in the
Universe is due to Charlier (1908, 1922), who followed earlier
concepts from Fournier d'Albe (1907). Nevertheless, the hierarchical
matter structuring concept is by no means new, but goes to as far back
as the pressocratic Greek philosopher Anaxagoras (Gruji\'c 2001), and
having in its list of defenders in the last 450 years thinkers like
Giordano Bruno, Kant, Leibniz (who opposed Newton's homogeneous universe
view) and Laplace, among others (see Gruji\'c [2001, 2002] and Baryshev
and Teerikorpi [2002] for historical accounts of those pre-modern
hierarchical ideas). More recently, however, the dispute between these
two views took an interesting new turn when Mandelbrot (1977, 1983)
and Pietronero (1987) proposed that the old hierarchical concept is
nothing more than the assumption that the matter in the Universe
should be distributed
according to a fractal pattern.\footnote{Wertz (1970, 1971) had reached
similar conclusions much earlier when he advanced a model mathematically
identical to Pietronero's. Although he never used the word ``fractal'',
Wertz's treatment did assume a scale-invariant, or self-similar, pattern
for the galaxy distribution (see Ribeiro [1994] and Ribeiro and
Miguelote [1998] for comparisons between these two models). Unfortunately
his model remained virtually unknown for decades, even to Mandelbrot
(1977, 1983) and Pietronero (1987).}

An important technical aspect of this dispute is that during most of its
modern century-old history, hierarchical modelling remained almost
exclusively confined to the realm of Newtonian cosmology (Charlier 1922;
Wertz 1970, 1971; Haggerty and Wertz 1972; Mandelbrot 1977, 1983;
Pietronero 1987; Coleman and Pietronero 1992; Sylos-Labini et al.\ 1998;
Ribeiro and Miguelote 1998; Abdalla et al.\ 1999; Pietronero and
Sylos-Labini 2000; Pietronero et al.\ 2003). Although
20th century scientific cosmology was basically relativistic cosmology
since its establishment in 1917 by Einstein, attempts to model the
hierarchical concept within a relativistic framework had to wait until
the 1970s when the first tentative relativistic models were advanced
(Bonnor 1972; Wesson 1978, 1979).\footnote{Earlier discussions on the
relativistic side of this dispute were reported by Wertz (1970; see
also Ribeiro 1994).}

In the early 1990s, this author took a fresh view of the relativistic
side of this issue, and concluded that when a fully relativistic model
is considered the problem of observationally identifying whether or not
a uniform distribution of matter is really being observed is much
less straightforward than previously assumed (Ribeiro 1992ab, 1993,
1994). Following this line of reasoning to its logical consequences,
Ribeiro (1995) concluded that in an Einstein-de Sitter (EdS) cosmology
the spatial homogeneity of this model may only be observed at much
closer ranges than previously assumed, up to $z \approx 0.01$, which
led this author to conjecture that the observed fractal structure may
be an observational effect of geometrical nature due to the fact that
astronomical observations are always carried out along our past null
cone. This means that the fractal, or hierarchical, hypothesis for the
matter distribution in the Universe may not be in contradiction with
the Cosmological Principle (Ribeiro 2001ab).

The conclusions reached in Ribeiro (1995) regarding the observable
depth of the homogeneous region in the EdS model were, however,
challenged by C\'el\'erier and Thieberger (2001) who claim that,
contrary to Ribeiro (1995), spacetime curvature will only show its
effects at redshift depths much larger than the range calculated
in Ribeiro (1995).

The aim of this paper is to discuss further the issues behind the
relativistic side of this dispute, as well as providing a reply to
C\'el\'erier and Thieberger (2001). This article also presents
results which extend and complement both Ribeiro (1995) and (2001b).
In what follows it is shown that the { radial (non-average) statistical
tools used in fractal analyses of the galaxy distribution, namely
the differential density $\gamma$ and the integral differential 
density $\gamma^\ast$, } depend strongly on the choice of observational
distance. Taking the EdS cosmology as the model of choice due to
its simplicity, it is possible to define four different observational
distances, namely the area distance $\da$, galaxy area distance $\dg$,
redshift distance $d_z$ and luminosity distance $\dl$. Proper and
comoving distances are also defined, but the former is equal to
$\da$ whereas the latter is equal to $\dg$ times a constant factor.
The distance given by Mattig's formula is also used, but it turns
out to be the same as $\dg$. In addition, it is shown that in view
of the fact that some of these observational distances go
asymptotically to infinity as one approaches the big bang whereas
others tend to zero, the above mentioned statistical functions are
strongly affected. In particular, the $\gamma$ statistic defined with
the area distance diverges at $z=1.25$, losing then its analytical
usefulness for higher redshifts.

The study presented here also suggests that the galaxy area distance,
the comoving distance and, still, the distance given by Mattig's
formula are unsuitable to probe for a fractal pattern as they
render both $\gamma$ and $\gamma^\ast$ constant for all redshift.
{ Therefore, if a fractal pattern really exists at ranges greater
than 100Mpc, which is the scale where both $\gamma$ and
$\gamma^\ast$ diverge depending on the choice of distance,}
this pattern could only be found at that scale if one carries out
statistical analyses with either the redshift or luminosity
distances, as these are the only distance definitions capable of
leading $\gamma$ and $\gamma^\ast$ to a decrease at higher
redshifts. Those findings confirm and complement similar results
obtained earlier by Ribeiro (1992b, 1995, 2001b).

{The dependence of the radial (non-average) statistical
functions on the choice of distance} can be seen if we remember
that any galaxy statistic requires us to define the area of a
spherical shell of a certain radius and since we have more than
one distance definition we will inevitably end up with various
expressions for such shell areas, none of them being better than
the others, a conclusion opposite to C\'el\'erier and Thieberger
(2001). Actually, their criticism confuses observables with the
relationships that relate them to one another and the underlying
cosmological model. Clarifying this point renders incorrect both
their calculations and interpretation. The main misconception is
about relativistic distances and how they are defined and used.
Inasmuch as this issue, cosmological distances, still causes
confusion, and is at the heart of this and related issues
(Ribeiro 2001b), further discussions of this subject is justified.
If the meaning and use of cosmological distances are clarified,
perhaps future errors of this sort will be avoided.

The plan of the paper is as follows. Sect.\ 2 reviews the issue of
measuring distances of cosmological sources. Sect.\ 3 develops
radial $\gamma$ type dust statistics in the EdS cosmology, and
{ section 4 analyzes the results and suggests that observers
should consider probing for a possible fractal pattern in the galaxy
distribution by using average statistics with the various distance
definitions used in this paper in order to ascertain whether or
not distortions due to the choice of distance are also present
in the final results.} The paper ends with a
concluding section.

\section{Measuring distances of cosmological sources}

In cosmology, observational distances are defined by the method of
measurement. This means that once we collect some astronomical data
of cosmological relevance, say, for instance, redshift of distant
galaxies, their apparent magnitudes, some intrinsic source size,
etc, the particular way we relate one observational quantity to
another defines a particular observational distance. In other words,
in cosmology, measuring a distance depends on circumstances. In a
flat and Euclidean space all methods of measurement lead to the
same and unique distance. In relativistic cosmology, however, that
is not the case and we are then forced to give up the notion of a
correct, or right, distance. In relativity, distance is not an
absolute concept.

The lack of a unique distance notion means that other quantities
used in cosmology where distance appears in their definitions must not be
unique either. This is the case of volume, spherical shell, average
density, and, of course, any statistic of galaxy distribution since
they require distance-derived quantities. However, as in our
neighbourhood we can take a Newtonian cosmology approximation where
all relativistic distances become the same, a relevant question then
arises as to where those quantities will start to deviate from these
Newtonian values. This question is far from trivial as general relativity
is a non-linear gravitational theory and this means that each
observational quantity may have its own particular deviation regime,
and which may substantially differ from other ones (Ribeiro 2001b).

Observational distance definitions in cosmology are an issue that has
been already well established in relativistic cosmology for over thirty
years since G.\ F.\ R.\  Ellis' (1971) seminal contribution to this subject.
In observational cosmology, however, the generality of Ellis' conclusions
do not yet seem to have been fully appreciated. For instance, Kayser,
Helbig and Schramm (1997) dealt with cosmological distances, as did 
Hogg (1999), who collected the various definitions in use in the
astronomical literature. Nevertheless, neither quoted
Ellis' (1971) fundamental work on this matter, nor did they seem to
have appreciated Ellis' results that showed how to calculate
cosmological distances for {\it any} cosmological spacetime metric.
Actually, the roots of this discussion go back as far as to the
famous Etherington reciprocity theorem, proved in 1933, and
rediscovered by R.\ Penrose and R.\ K.\ Sachs in the 1960s (Ellis
1971, p.\ 111; see also Schneider, Ehlers and Falco 1992).

An observer, of course, may wisely ignore this whole discussion if
one is only interested in relating observational quantities to
other observational quantities. In this approach, which may be called
``Sandage's prescription'' (see Ribeiro 2001b), distances become an
internal parameter of the theory and the choice of distance turns out
to be irrelevant. The limitation of this approach is that if the
final quantities are explicitly dependent on distances, which is the
case for the average density and many statistical tools of galaxy
distribution, Sandage's prescription is inapplicable.

Having said all above, it is therefore wise to approach the subject
theme of this paper very carefully. Thus, I start by describing
the various observational distances appearing in Ellis' (1971) paper,
and then proceed to the theoretical issues.

\subsection{Observational distances}

The {\it luminosity distance} $\dl$ is defined if one assumes a flat
and non-expanding universe, as if it were Euclidean, and relates apparent
magnitude, or observed flux $F$, with the {\it intrinsic} luminosity $L$
of the source by means of the well-known expression
\be F=\frac{L}{4\pi { \left( \dl \right) } ^2}. \lb{dl} \ee

The {\it area distance} $\da$ (also known as {\it angular diameter
distance}, {\it corrected luminosity distance} or {\it observer
area distance})\footnote{In this paper I will adopt Ellis' (1971)
terminology for observational distances, as I believe his name
choices to be the least confusing, and, geometrically, most
appropriate.} is measured by means of the relationship between an
{\it intrinsically measured} cross-sectional area element $\dsa$
of the source and the observed solid angle $\doa$ (see figure
\ref{figda}),
\be \dsa = \doa { \left( \da \right) }^2. \lb{da} \ee

The {\it galaxy area distance} $\dg$ (also known as {\it 
effective distance}, {\it angular size distance}, {\it transverse
comoving distance} or {\it proper motion distance}) is similar,
but defined from the opposite point where the area distance is
specified. The solid angle $\dog$ is measured at the rest frame of the
galaxy, whereas the intrinsic cross-sectional area $\dsg$ is
measured here (see figure \ref{figdg}). The defining equation may be
written as,
\be \dsg = \dog { \left( \dg \right) }^2. \lb{dg} \ee

The methods above for relating observables define distances
that are in principle directly measurable and do not
necessarily need to  coincide.\footnote{By means of galactic parallax
one can still define and measure a fourth distance, the
{\it parallax distance} $d_{\ssty P}$ (Ellis 1971). However, we are
still unable to measure galactic parallaxes and, therefore,
$d_{\ssty P}$ is of no use in our discussion here.} Besides $\dl$, $\da$,
$\dg$, the equations above also define three quantities which are
measured here, that is, at the observer's location, namely $F$,
$\doa$, $\dsg$, and, additionally, three other quantities which
can only be measured at the source's rest frame, namely $L$,
$\dsa$, $\dog$.

Notice that as $\dl$ and $\da$ are directly determinable quantities,
without assuming any cosmological model, their measurements are,
therefore, model independent. In fact, precise determination of these
quantities can be used to distinguish among cosmological models.
$\dg$ is not directly measurable since $\dog$ cannot be determined
from the observer's location. To be useful, $\dg$ requires a further
equation relating it to $\dl$ or $\da$ (see \S \ref{rt} below).

\subsection{Etherington's reciprocity theorem}\lb{rt}

The observational distances discussed above are connected to one
another and the redshift $z$ by a very important theorem proved a
long time ago by Etherington (1933), called the {\it reciprocity
theorem} or {\it Etherington's reciprocity law}. It may be written
as (Ellis 1971; Schneider, Ehlers and Falco 1992, p. 111, 116),
\be \dl = { \left( 1+z \right) }^2 \da = \left( 1+z \right) \dg,
    \lb{rec}
\ee
and it is valid for {\it any} cosmological model. In fact the theorem
is purely geometrical and only requires that source and observer
are connected by null geodesics. It is simply a consequence of the
geodesic deviation equation and it tells us that if $z=0$ that
implies that equal surface elements $\dsg$, $\dsa$ subtend equal
solid angles $\dog$, $\doa$, irrespective of the curvature of
spacetime (Ellis 1971). As this theorem is valid for all cosmologies,
it has been recently suggested that it can be a powerful tool
for testing non-standard cosmological models, as well as being capable
of distinguishing between various models of dark energy by 
comparing distances measured by standard candles, that is, $\dl$,
and standard intrinsic dimensions, i.e., $\da$ (Basset and Kunz 2004;
Uzan et al.\ 2004).

\subsection{The necessity of a cosmological model}

So far we have not needed to establish any cosmological model,
but as we have five expressions (eqs.\ \ref{dl}, \ref{da},
\ref{dg}, \ref{rec}) containing four quantities measurable here,
$F$, $\doa$, $\dsg$, $z$, plus six unknown quantities, $L$, $\dl$,
$\dsa$, $\da$, $\dog$, $\dg$, we end up with an unsolvable system.
However, if we are able to independently determine either $L$ or
$\dsa$, the system can be solved. In both cases we require knowledge of
the intrinsic physics of the source, or source evolution, a theory
which is still lacking for galaxies.

There is a possible way to solve the system by determining $\da$.
Let us call $\alpha$ the observed angular dimension of some object
whose linear {\it intrinsic} dimension $\ella$, perpendicular to the
line of sight can be estimated (see figure \ref{figda}). Then,
\be \ella \approx \da \alpha, \lb{la} \ee 
and we can determine the area distance to a reasonable degree of
accuracy. Notice, however, that $\ella$ is measured {\it at the
source's rest frame}, which means that we again need knowledge of
the intrinsic physics of the source in order to estimate $\ella$
and then calculate $\da$. This remark will be very important in
what follows.

Another attempt at looking at the difficulties above was the
{\it ideal observational cosmology program}, which proposed to
characterize in detail the way in which cosmological observations
can be directly used to determine the cosmological spacetime geometry
(Ellis et al. 1985). The main motivation behind this program
is to determine the spacetime metric of our universe directly
from astronomical observations, without assuming a cosmological model
beforehand. In the process of doing that, one must first determine
what is and what is not decidable in cosmology on the basis of
astronomical observations. Although this is an appealing idea, this
program cannot be fully implemented due to the astrophysical evolution,
both in number and in luminosity, of galaxies. To implement this
program we require precise observations of $\dl$ or $\da$, redshift,
galaxy number counts and an adequate model of how the populations
and luminosities of galaxies of different morphologies evolve with
$z$, that is, to know independently the function $L=L(z)$ for each
morphological galaxy type. It is presently impossible to obtain all
that information without assuming a cosmological model.

Another way of possibly solving this system of equations and
determining all quantities described so far is to find
a homogeneous class of sources, that is, objects which share some
basic features, and whose intrinsic luminosity $L$ may be known.
This is, of course, the idea behind supernova cosmology. The
limitations of this approach are that the measured intrinsic
supernova Ia luminosities at high $z$ assume no luminosity
evolution at different redshifts due to limitations in our
theoretical understanding of these objects (Perlmutter and Schmidt
2003; Li and Filippenko 2003). In addition, the building blocks
of the universe are galaxies and not supernovas, and the picture
emerging from supernova cosmology ought to be compatible with
galactic physics and observations. Nevertheless, this is currently
a very promising line of research in observational cosmology.

So, unless supernova cosmology, or another observational methodology,
firmly establishes cosmological model-independent relationships between
values of $\dl$ and $\da$ at different $z$, we have no other way but to
assume an universe model in order to obtain another equation, usually
in the functional form of either $\dl(z)$, or $\da(z)$, or $\dg(z)$,
and be able to fully solve the system of equations above.

\subsection{Theoretical distances}

Once a cosmological model is assumed, we can, of course, define
other, {\it model dependent}, distances like the comoving distance,
proper distance, interval distance, 
geodesic distance, absolute distance, etc. These are all,
however, different forms of line element separations ($ds^2$),
whose theoretically-defined expressions are entirely dependent on
the spacetime geometry {\it and} the particular solution of Einstein's
field equations.\footnote{Notice that the reciprocity theorem
does not require a solution of Einstein's field equations, but only
the pseudo-Riemannian spacetime.} Their particular expressions
change according to the cosmological model and, hence, they are
{\it not} observationally defined distances, although they do play
an important theoretical role in cosmology.

\subsection{The consistency problem in observational
            cosmology}\lb{consistency}

Since most observational results in cosmology either explicitly or
implicitly assume a cosmological model, usually the standard one, 
in their presentation, a question arises immediately. Do the
observational results really determine parameters compatible with
the assumed cosmological model, as they should? This is the
{\it consistency problem} in observational cosmology (Ribeiro and
Stoeger 2003). Consistency tests are, therefore, desirable, in
order to check the agreement between the derived observational
parameters and the cosmological model they assume. This means 
that we can only discuss the possible departures from a
homogeneous matter distribution if we choose an average density
that theoretically allows such a departure and then check the model
for consistency with the data. In order words, once we calculate
observational parameters by means of a cosmological model, the
agreement or disagreement between the data and the model will be
tested by consistency or inconsistency.

\section{Radial dust statistics in the Einstein-de Sitter cosmology}

Let us define another relation which may be called an
observational distance, the {\it redshift distance} $d_z$. It
may be written as,
\be d_z=\frac{cz}{H_0}, \lb{dz} \ee
where, $c$ is the light speed and $H_0$ is the Hubble constant. 
This is, of course, just a consequence of the velocity-distance
equation and the Doppler approximation in a expanding universe,
being valid only for $z<1$ (Harrison 1993). It is not an
observational distance in the sense as discussed above, but a
quantity proportional to small redshifts in the standard cosmology.
However, since this is the defining equation of the Hubble law
(Harrison 1993) and is often used in observational cosmology, 
it is useful to list it here alongside the previously discussed
cosmological distances and adopt equation (\ref{dz}) as the
{\it defining expression} of $d_z$ for all $z$.

\subsection{Radial (non-average) statistical tools}\lb{radial}

Let us now call generically by $d_0$ some observationally defined
distance. It can be any of the four distances $\dl$, $\da$, $\dg$,
$d_z$ defined so far. If $S_0$ is the area of the observed spherical
shell of radius $d_0$, and $V_0$ the observed volume of radius $d_0$,
we have that,
\be S_0= 4 \pi {\left( d_0 \right) }^2, \lb{s0} \ee
\be V_0= \frac{4}{3} \pi {\left( d_0 \right) }^3. \lb{v0} \ee

{Following J.\ Wertz, I define the {\it differential
density} $\gamma(d_0)=\gamma_0$ at a certain distance $d_0$ as being
given by the following expression (Wertz 1970, 1971; see also Ribeiro
and Miguelote 1998),}
\be \gamma_0 = \frac{1}{S_0} \frac{dN_c}{d(d_0)}, \lb{gama} \ee
where $N_c$ is the {\it cumulative radial number count from the origin}
of sources.

{The {\it integral differential density}
$\gamma^\ast(d_0)=\gamma^\ast_0$ is the integration of
$\gamma_0$ over the observational volume $V_0$.} Its equation
yields
\be \gamma^\ast_0 = \frac{1}{V_0} \int_{V_0} \gamma(d_0) dV_0
    =\frac{3}{{\left( d_0 \right) }^3} \int_0^{d_0} x^2 \; \gamma(x) dx.
    \lb{gamas}
\ee

{The two definitions above look similar to the conditional densities
$\Gamma$ and $\Gamma^\ast$ appearing in Pietronero (1987). However,
it is important to point out the difference between the radial
(not averaged) $\gamma$ and $\gamma^\ast$ statistics and the
non-radial, but averaged, $\Gamma$ and $\Gamma^\ast$ statistics.
The differential density $\gamma$ and its integral} version measure
how the density scales as a function of distance, whatever distance
choice, whereas the conditional density $\Gamma$ is an average
quantity, measuring the density from an occupied point and
then averaged over all realizations of a given stochastic process.
{ For the galaxy distribution what one does is to use a volume
average to calculate the conditional density $\Gamma$ and the average
conditional density $\Gamma^\ast$, which is an averaged integral of
$\Gamma$ (Pietronero 1987; Coleman and Pietronero 1992; Sylos-Labini
et al.\ 1998; Pietronero 2003). Therefore, computing radial ($\gamma$)
and non-radial ($\Gamma$) statistics are the result of 
different operations.}

Therefore, here I follow Wertz's (1970, 1971) original
contribution and will be approaching this problem from a regular
(analytical) viewpoint, in addition to assuming that classical
general relativity is a valid and appropriate tool for modelling
the smoothed out large scale distribution of galaxies.
{ This means that this paper deals with radial (non-average)
$\gamma$ statistics derived from theory and attempts to see what the
final results could suggest for the calculation of the non-radial
(average) $\Gamma$ statistics obtained from galaxy
catalogues.\footnote{ I have not used Wertz's original
symbol $\rho_{\ssty D}$ for the differential density as it could be
confusing in the present context.}}

\subsection{Radial statistics of the Einstein-de Sitter cosmological model}

The EdS metric may be written as (from now on $c=G=1$),
\be ds^2=dt^2-a^2(t) \left[ dr^2 + r^2 \left( d\theta^2 + \sin^2
    \theta d \phi^2 \right) \right]. \lb{eds}
\ee
The observational distances will then be given by the expressions below,
\be \dl(z)=\frac{2}{H_0} \left( 1+z- \sqrt{1+z} \right), \lb{dlz} \ee
\be \da(z)=\frac{2}{H_0} \frac{\left( 1+z- \sqrt{1+z} \right)}{{ \left(
            1+z \right) }^2},
    \lb{daz}
\ee
\be \dg(z)= \frac{2}{H_0} \left( \frac{ 1+z- \sqrt{1+z}}{1+z} \right),
    \lb{dgz}
\ee
whereas the cumulative number count yields,
\be N_c(z)= \frac{4}{H_0 M_g} { \left( \frac{ 1+z- \sqrt{1+z}}{1+z}
            \right) }^3. 
    \lb{ncz}
\ee
where $M_g$ is the average galactic rest mass ($\sim 10^{11}M_\odot$).

Considering equations (\ref{s0}) and (\ref{v0}) we can define four
observational spherical shells and volumes as follows: 
\be S_{\ssty L} = \frac{16 \pi}{{(H_0)}^2} { \left( 1+z-\sqrt{1+z}
                  \right) }^2,
    \lb{slz}
\ee
\be S_{\ssty A} = \frac{16 \pi}{{(H_0)}^2} \frac{{ \left( 1+z-\sqrt{1+z}
                  \right) }^2}{{ \left( 1+z \right) }^4},
    \lb{saz}
\ee
\be S_{\ssty G} = \frac{16 \pi}{{(H_0)}^2} { \left( \frac{1+z-\sqrt{1+z}}{1
                  +z } \right) }^2,
    \lb{sgz}
\ee
\be S_z = \frac{4 \pi z^2}{{(H_0)}^2}, \lb{szz} \ee
\be V_{\ssty L} = \frac{32\pi}{3(H_0)^3}{ \left( 1+z-\sqrt{1+z} \right) }^3,  
    \lb{vlz}
\ee
\be V_{\ssty A} = \frac{32\pi}{3(H_0)^3} \frac{{ \left( 1+z-\sqrt{1+z}
                  \right) }^3}{{ \left( 1+z \right) }^6},
    \lb{vaz}
\ee
\be V_{\ssty G} = \frac{32\pi}{3(H_0)^3} { \left( \frac{1+z-\sqrt{1+z}}{1
                  +z } \right) }^3, \lb{vgz} \ee
\be V_z = \frac{4 \pi z^3}{3{(H_0)}^3}. \lb{vzz} \ee

It is convenient to express all quantities in terms of the redshift.
Therefore, equation (\ref{gama}) should be rewritten as below,
\begin{eqnarray}
\gamma_0 & = & \frac{dN_c}{dz} {\left[ S_0 \frac{d}{dz}(d_0)
    \right] }^{-1} \nonumber \\
    & = & \frac{6}{H_0M_g} \frac{ {\left( 1+z- \sqrt{1+z}
    \right) }^2}{{ \left( 1+z \right) }^{7/2}} {\left[ S_0
    \frac{d}{dz}(d_0) \right] }^{-1}  .
    \lb{gama2}
\end{eqnarray}
We are now able to write the differential density in terms of the
redshift for each observational distance that enters in its defining
equation above. They yield,
\be \gamma_{\ssty L} = \frac{\mu_0}{ \left(
    2 \sqrt{1+z} - 1 \right) { \left( 1+z \right) }^3},
    \lb{gl}
\ee
\be \gamma_{\ssty A} = \frac{ \mu_0 { \left(
    1+z \right) }^3}{\left( 3 -2 \sqrt{1+z} \right) },
    \lb{ga}
\ee
\be \gamma_{\ssty G} = \mu_0,
    \lb{gg}
\ee
\be \gamma_z = \frac{4 \mu_0 { \left( 1+z-
    \sqrt{1+z} \right) }^2}{z^2 { \left( 1+z \right) }^{7/2}},
    \lb{gz}
\ee
where,
\be \mu_0 = \frac{3 {(H_0)}^2}{8\pi M_g}. \lb{mu0} \ee

Two conclusions can immediately be drawn from the equations above. Firstly,
the differential density $\gamma_{\ssty G}$, calculated by means of the
galaxy area distance $\dg$, is constant for all redshift. In other words,
the use of such a distance makes the analysis completely insensitive to
any change in the dust distribution of the EdS cosmology. In other words,
using $\dg$ we reach the conclusion that the dust distribution is
homogeneous everywhere, even along the past null cone (or the lookback
time). This is a built-in feature of the model and means that this
quantity is of no use for probing the possible inhomogeneity of the
observable Universe. Such a result was also reached by Ribeiro (2001b)
by means of a somewhat different path. Secondly, it is clear from
equation (\ref{ga}) that $\gamma_{\ssty A}$ diverges at $z=1.25$,
losing then its analytical usefulness for higher redshifts.

Finally, we are now in a position to calculate the integral differential
density for each observational distance. If we rewrite equation
(\ref{gamas}) in terms of the redshift we get that,
\be \gamma^\ast_0 (z)=\frac{1}{V_0(z)} \int_0^z \gamma_0 (z)
    \frac{dV_0(z)}{dz} \; dz.
    \lb{gamas0z}
\ee
Using the expressions obtained above for volume and the differential
density in each observational distance we obtain the following equations
for the integral differential density,
\be \gamma_{\ssty L}^\ast = \mu_0 { \left( 1+z \right) }^{-3},
    \lb{gamasl}
\ee
\be \gamma_{\ssty A}^\ast = \mu_0 { \left( 1+z
    \right) }^{3},
    \lb{gamasa}
\ee
\be \gamma_{\ssty G}^\ast = \mu_0,
    \lb{gamasg}
\ee
\be \gamma_z^\ast = 8 \mu_0 { \left[ \frac{ 1+z-\sqrt{1+z}
                    }{z { \left( 1+z \right) }} \right] }^3.
    \lb{gamasz}
\ee

It is useful to define the {\it average numerical density}
$\langle n \rangle$ of a certain observational distance $d_0$ as
being given by 
\be \langle n_0 \rangle = N_c/V_0. \lb{n0} \ee
With the expressions obtained so far it is straightforward to show that
the equations below hold:
\be \langle n_{\ssty L} \rangle = \gamma_{\ssty L}^\ast,
    \lb{nl}
\ee
\be \langle n_{\ssty A} \rangle = \gamma_{\ssty A}^\ast,
    \lb{na}
\ee
\be \langle n_{\ssty G} \rangle = \gamma_{\ssty G}^\ast,
    \lb{ng}
\ee
\be \langle n_z \rangle = \gamma_z^\ast.
    \lb{nz}
\ee

All results above agree and expand both Ribeiro (1995) and (2001b).
The former, however, limited the analysis to the luminosity distance,
while the latter concentrated on the behaviour of the average
densities and their implications for the possible observational
smoothness of the Universe. Equations (\ref{gl}) and (\ref{gamasl})
are equal to the ones appearing in Ribeiro (1995), with the difference
that here they are a function of the redshift, whereas in Ribeiro
(1995) they were written in terms of $\dl$.

Notice again that $ \gamma_{\ssty G}^\ast$ and, therefore,
$ \langle n_{\ssty G} \rangle $ are constant, meaning that if one
carries out consistency checks (see \S \ref{consistency} above)
between possible observable departures from homogeneous distribution,
even at small scales ($z<0.1$) one should not use the galaxy area
distance $\dg$. It would seem incorrect to do so because this distance
renders constant its associated average density for all $z$.

\subsection{Theoretical distances}\lb{tdis}

We should now relate the expressions obtained above with other
theoretical distances which often appear in the literature. The
{\it proper volume} is defined as follows,
\be dV_{\ssty PR} =a^3 r^2 dr \sin \theta d \theta d \phi,
    \lb{dvpr}
\ee
and the {\it comoving volume} yields,
\be dV_{\ssty C}=r^2 dr \sin \theta d \theta d \phi.
    \lb{dvc}
\ee
It is easy to show that in the EdS cosmology the area distance and the
{\it proper distance} $d_{\ssty PR}$ are given by  {\it the same}
expression (Ribeiro 1992b, 1995, 2001b; Ribeiro and Stoeger 2003),
\be d_{\ssty PR} = a[t(r)] \; r =\da, \lb{dpr} \ee
and, therefore, have the same relationship to the redshift as given
by equation (\ref{daz}) above.

As it is well known, the {\it comoving distance} $d_{\ssty C}$ may
be written as below,
\be d_{\ssty C}=r=\frac{d_{\ssty PR}}{a}= { \left( \frac{18}{H_0}
    \right) }^{1/3} \left( \frac{ 1+z- \sqrt{1+z}}{1+z} \right), 
    \lb{dc}
\ee
since
\be a= { \left[ \frac{4}{9 (H_0)^2} \right] }^{1/3} \frac{1}{1
       +z}. 
       \lb{az}
\ee
Therefore, comparing with equation (\ref{dgz}) it follows that,
\be \dg = { \left[ \frac{4}{9 (H_0)^2} \right] }^{1/3} d_{\ssty C}.
    \lb{dgdc}
\ee
Thus, the comoving distance, and, as a consequence, the comoving
volume, are just the galaxy area distance times a constant factor,
having then the same analytical behaviour with $z$ as any expression
containing $\dg$.

It is often the case in studies of galaxy statistics to use the famous
Mattig formula to convert redshifts to distances. In the EdS cosmology it
is easily proven that 
\be d_{\ssty \rm MATTIG} = \frac{2}{H_0} \left( \frac{ 1+z-
    \sqrt{1+z}}{1+z} \right) = \dg. 
    \lb{mattig}
\ee
So, the distance given by Mattig's formula is nothing more than the
galaxy area distance.

The results obtained above show that the analysis presented in this
paper does not need to refer to these often adopted theoretical distances,
since by studying the behaviour of $\da$ and $\dg$ we will have them
taken into account.

\subsection{Reply to C\'el\'erier and Thieberger (2001)} 

In attempting to justify their different approach to the problem,
C\'el\'erier and Thieberger (2001; from now on CT01) have stated
the following on page 452 of their paper: ``... the luminosity
distance is the observable quantity relevant for {\it radially}
measured distances (...) But, when looking at a cross sectional
area (...) perpendicular to the light ray and subtending a solid
angle (...), the observer must consider the area distance.'' Then
they obtain expressions for the radial differential density and
volume element different from the ones appearing above. Those
expressions were then used as the basis for their subsequent
study and conclusions.

CT01 seem to have assumed that because the area distance $\da$
is defined by means of a cross sectional area $\dsa$, which is
measured at the source's rest frame, it is then the ``correct''
quantity to define the area of the observed spherical shell,
with all other distances being, therefore, incorrect to use in
this case. Moreover, they believe that $\dl$ is the ``correct''
distance for radial measure. Let us see why this reasoning is
not correct. 

Any observed distance, be it $\dl$, $\da$, $\dg$ or $dz$, define an
observable spherical shell and a radial measure for a given $z$.
Because these distances are different, the shells will then have
different radii and areas, but, nevertheless, they can be calculated
from observations because these distances can be defined and observed.
Even theoretically defined distances like the comoving and proper
distances will also define spherical shells. The particular
aspect about the area distance is that {\it if}, and {\it only if},
in our astronomical observations we, {\it independently of a cosmological
model}, had also been able to observe and tabulate {\it intrinsically
measured dimensions} such as $\dsa$ or $\ella$ (see eqs.\ \ref{da},
\ref{la} and figure \ref{figda}), then the method to relate those
intrinsically measured dimensions with the observed solid angle
$\doa$, or the observed angular dimension $\alpha$, would be by
means of equations which will result in measuring the area
distance $\da$. This has nothing to do with the definition of
spherical shells and measuring their areas or radius. 

Similarly, if, independently of a cosmological model, we were able to
measure the intrinsic luminosity $L$ of a cosmological source, then
by measuring its observed flux $F$ we will be able to relate
these two quantities and obtain a distance called luminosity distance.
Such a procedure has nothing to do with a so-called ``radial distance
measure'', but it is simply a method to obtain a specific distance
by means of two different, but directly observable, quantities.
Supernova cosmology is, of course, based exactly on this methodology
and, due to that, it is able to actually verify a cosmological model
rather than assume one (Perlmutter and Schmidt 2003).

CT01 confuse two different distances when they wrote their
differential density.\footnote{ Both Ribeiro (1995) and CT01
calculated radial (non-average) $\gamma$ statistics, but named them
with the symbol $\Gamma$ for non-radial (average) statistics. As
discussed above (see \S \ref{radial}) $\gamma$ and $\Gamma$ are
different quantities and, therefore, one should use different symbols
to avoid confusion.} With the notation adopted in this paper CT01's
equation (20) can be written as follows:
\begin{eqnarray}
\gamma_{\ssty \rm CT01} & = & \frac{1}{S_{\ssty A}} \frac{dN_c}{d(\dl)}
    = \frac{dN_c}{dz} {\left[ S_{\ssty A} \frac{d}{dz}(\dl)
    \right] }^{-1} \nonumber \\
    & = & \frac{3 {(H_0)}^2}{8\pi M_g} \frac{ \left( 1+z
    \right) }{ \left( 2 \sqrt{1+z} - 1 \right) }.
   \lb{gamact01}
\end{eqnarray}
Notice how two entirely different methods for measuring the distance
of a cosmological source are included in the same equation. They
further defined an observed volume element as follows (CT01's
eq.\ 21),
\be dV_{\ssty \rm CT01} = S_{\ssty A} \; d(\dl).
    \lb{dvct01}
\ee
Everything else in their paper follows from these two equations.  
It is important to point out that as at the same redshift
$\dl \not= \da$ with $\dl > \da$ (see \S \ref{rt} above),
in both expressions above there will be a mismatch between
$S_{\ssty A}$ and $\dl$. So, for a given $z$ the volume
element $dV_{\ssty \rm CT01}$ will have a spherical area
$S_{\ssty A}$ whose radius $\da$ is smaller than the
distance $\dl$ where this spherical area is supposed to be
located. It seems difficult to justify such a procedure.

As we shall see below, the results obtained by CT01 are different from
Ribeiro's (1995) because densities defined with the area distance
grow forever, inasmuch as $\da$ has the peculiar feature of
vanishing at the big bang. However, the luminosity distance behaves
in an opposite manner and goes to infinity at the big bang. Therefore,
a density defined with $\dl$ vanishes at the big bang, while
another density defined with $\da$ tends to infinity (see 
Ribeiro 2001b). By confusing these two distances CT01 smooth out
the eventual decrease of density in their expression, as compared
to Ribeiro's (1995), due to an incorrect algebraic manipulation.

\section{Analysis}

We are now in a position to discuss the results obtained in the previous
section. Figure \ref{fig-gama} shows graphs of the differential density
$\gamma$ for all observational distances (proper and comoving distance
behave as $\da$ and $\dg$ respectively -- see \S \ref{tdis} above). The
plots show very clearly the entirely different behaviour of
$\gamma_{\ssty A}$ on one hand, and $\gamma_z$ and $\gamma_{\ssty L}$
on the other. Notice that $\gamma_{\ssty G}$ is constant for all $z$
and that $\gamma_{\ssty A}$ diverges at $z=1.25$. Figure \ref{fig-gama-s}
shows the integral differential density $\gamma^\ast$ for each
observational distance. Notice again the entirely different behaviour
one obtains if one uses $\dg$ instead of $d_z$ or $\dl$.

The homogeneous behaviour in this case is given by the function obtained
with the galaxy area distance $\dg$, and the plots show quite clearly
that deviations from homogeneity start to occur at $z=0.01$. At $z=0.1$
this deviation is meaningful. Notice that if one adopts the distance
as given by Mattig's formula or the comoving distance one ends up with
the same behaviour as given by $\dg$, having therefore, no deviation
from homogeneity, {\it by construction}. 

These results {\it suggest} that one should {\it not}
adopt these distances when trying to probe whether or not the
galaxy redshift survey data show, or do not show, an eventual
homogeneous pattern. In this respect, it is important to point
out that many authors do adopt those apparently unsuitable
distances when trying to ascertain a possible homogenization of
the galaxy distribution. Among a few recent analysis, both
Tikhonov et al.\ (2000) and Mart\'{\i}nez et al.\ (2001) started
with Mattig's equation, while Cappi et al.\ (1998) adopted the
comoving distance. On the other hand Sylos-Labini et al.\ (1998),
Joyce et al.\ (1999) and Pietronero et al.\ (2003) seemed to
have used $d_z$ instead.

{ So, in view of the results presented here one may be compelled
to ask how those results may affect the statistics of data stemming
from galaxy redshift surveys.}

One is faced with a { fundamental difficulty if one tries
to answer this
question by means of the analysis presented above. The main
problem stems from the fact that the data gathered from
galaxy redshift surveys is presented in terms of non-radial
and averaged statistics, whereas here all quantities were
discussed in terms of radial and non-averaged ones.} As seen
above, radial and non-radial statistics are the result of
different operations (see \S \ref{radial}). As clearly shown
in figures \ref{fig-gama} and \ref{fig-gama-s} the use of
different choices of distance produce distortions in the radial
statistics and, thus, it is reasonable to suppose that similar
distortions could affect the non-radial statistics as well.
Nevertheless, there are no studies about this possible effect
in statistics derived from galaxy catalogues as the distance
problem has been mostly ignored so far in observational
cosmology by both sides of the fractal controversy (Joyce et al.\
[1999] have, nonetheless, showed that non-radial densities are
strongly affected by the kind of distance-redshift used).
Therefore, the presence and strength of such a possible
distortion showing itself in data obtained from galaxy catalogues
remains, so far, an open problem. 

Another important caveat is that the analysis presented above is
bolometric, while real data analysis stemming from redshift surveys
are always in a limited frequency range and often require
the use of some form of K-correction. As discussed elsewhere (Ribeiro
2002; Ribeiro and Stoeger 2003) a full relativistic analysis of
radial number count where those effects are taken into consideration
may affect very strongly the final quantities. In particular the
luminosity function can only be compared with theoretical number
count by means of complex equations which may seriously affect and
change the theoretical quantities (Ribeiro and Stoeger 2003). As
noticed by Joyce et al.\ (1999), a simple luminosity function where
those effects are not taken into account may change completely the
final results of whether or not the data from the ESP galaxy survey
show, or do not show, a scale invariant pattern. Other factors like
the type of FLRW model as well as the existence or not of dark
energy and a non-zero cosmological constant may also affect the
statistics in unpredictable ways. Finally, even the calculation of the
average density is affected by those effects (Ribeiro and Stoeger 2003).
It is beyond the scope of this paper to carry out a fully relativistic
analysis to derive the average density from the luminosity function
parameters, which themselves were calculated from galaxy
catalogues, that is, where the data was obtained with limited frequency
range observations, as well as some kind of morphological classification.
Such an analysis should shed some light on those issues (Albani and
Ribeiro 2004, in preparation), but we may well
assume in advance that those effects may be strong.

{ Considering those caveats the safest thing that could be said
at this stage is that observers should take into consideration the
possible distortion of their statistical analyses of galaxy surveys
due to the adoption of different distance definitions. This may be
especially true in the fractal controversy. Comparing data reduced with
the comoving distance or by means of the Mattig formula with data
reduced with the redshift distance could, perhaps, be one of the
sources of controversy. Therefore, it would be very interesting to
see the results if Cappi et al.\ (1998), Tikhonov et al.\ (2000)
and Mart\'{\i}nez et al.\ (2001) were to reduce their data by
assuming either $\dl$ or $d_z$, and if Sylos-Labini et al.\ (1998)
and Pietronero et al.\ (2003) were to do the same, but adopting
now $d_{\ssty C}$ or $\dg$. }

\section{Conclusion}

In this paper, I have discussed the effect of the distance choice in
statistical tools used for possible fractal characterization of the
galaxy distribution. I reviewed the basic notions of measuring
distances of cosmological sources and used the area distance $\da$,
the galaxy area distance $\dg$, the luminosity distance $\dl$, the
redshift distance $d_z$, as well as the comoving and proper distances,
$d_{\ssty C}$ and $d_{\ssty PR}$ respectively, to calculate the
differential density $\gamma$ and the integral differential
density $\gamma^\ast$ of the dust distribution in a
Einstein-de Sitter cosmology. The results showed the fundamental
role played by the choice of distance in the determination of the
scale where relativistic corrections must be taken into account
as both $\gamma$ and $\gamma^\ast$ are strongly affected by the
choice of distance. It is also shown that an inappropriate distance
choice may lead to a failure to find evidence of a galaxy
fractal structure when one calculates those two statistics, even
if a self-similar structure does exist in the galaxy distribution.
In particular, $\dg$, $d_{\ssty C}$ and the distance given by
Mattig's formula seem unsuitable to probe for a fractal pattern as
these two radial densities become constant for all redshifts when
those distances are used in their calculation. This is a built-in
feature of the EdS cosmological model. The analysis also suggests
that if a fractal pattern really exists at ranges greater than
100Mpc, it can only be detected at that scale by these
statistics if they are calculated with either $\dl$ or $d_z$, as
these two distance definitions are the only ones where, by
theory, $\gamma$ and $\gamma^\ast$ can decrease at higher redshifts.
I have also shown that C\'el\'erier and Thieberger's (2001) critique
of Ribeiro's (1995) earlier study, partially re-obtained here, is
incorrect due to misconceptions regarding relativistic distance
definitions, rendering their objections impaired.

\begin{acknowledgements}
Thanks go to M.\ A.\ H.\ MacCallum for discussions on cosmological
distances. I am also grateful to the referee for very useful remarks
which improved the paper. Partial support from FAPERJ and Instituto do
Mil\^enio CNPq 620053/2001-1 is acknowledged.
\end{acknowledgements}

\begin{figure*}
      \centering
      \includegraphics[width=17cm]{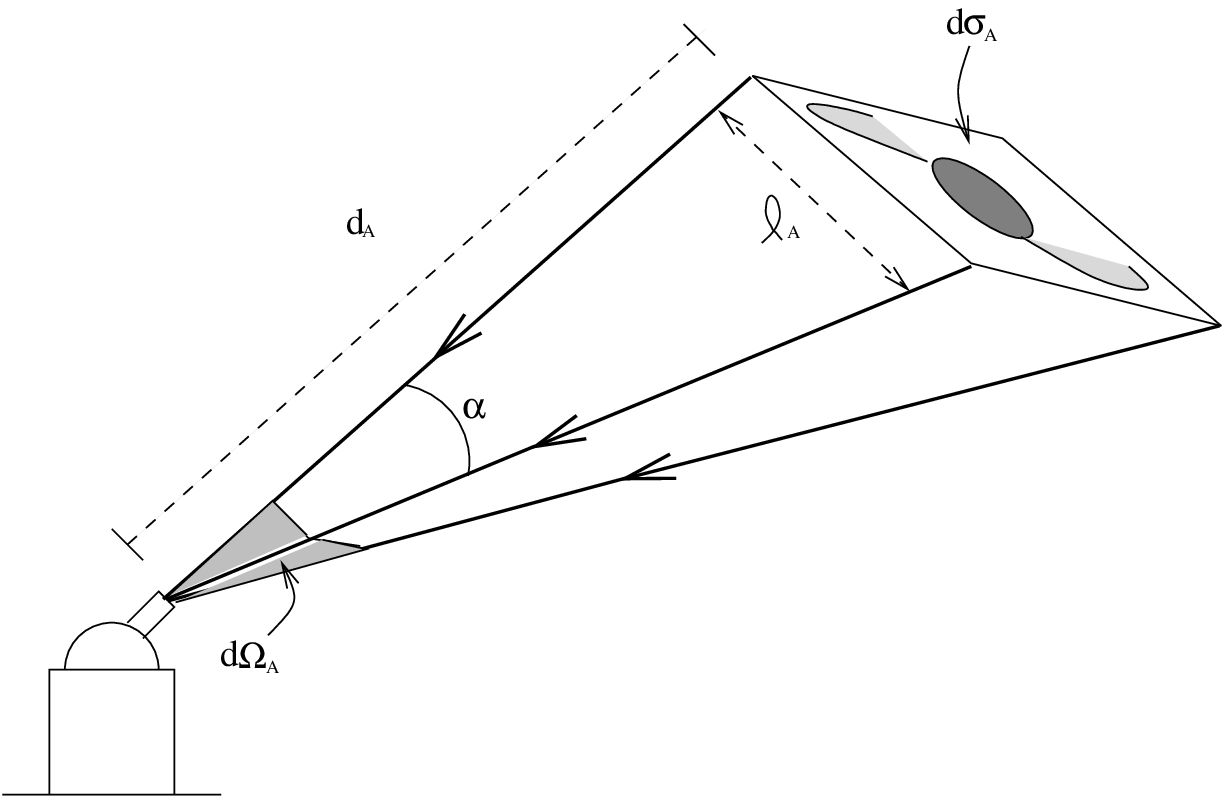}
      \caption{The area distance $\da$ is obtained by the relationship
               between the intrinsic cross-sectional area $\dsa$ of the
	       source, measured at its rest-frame, and the solid angle
	       $\doa$ measured by the observer (eq.\ \ref{da}). If an
	       intrinsic dimension $\ella$ can be measured at the source's
	       rest-frame and its corresponding angular dimension
	       $\alpha$ is measured by the observer, one can estimate
	       $\da$ (eq.\ \ref{la}). In the Einstein-de Sitter cosmology
	       $\da$ is the same as the proper distance (see
	       \S \ref{tdis}).}
      \lb{figda}
\end{figure*}
\begin{figure*}
      \centering
      \includegraphics[width=17cm]{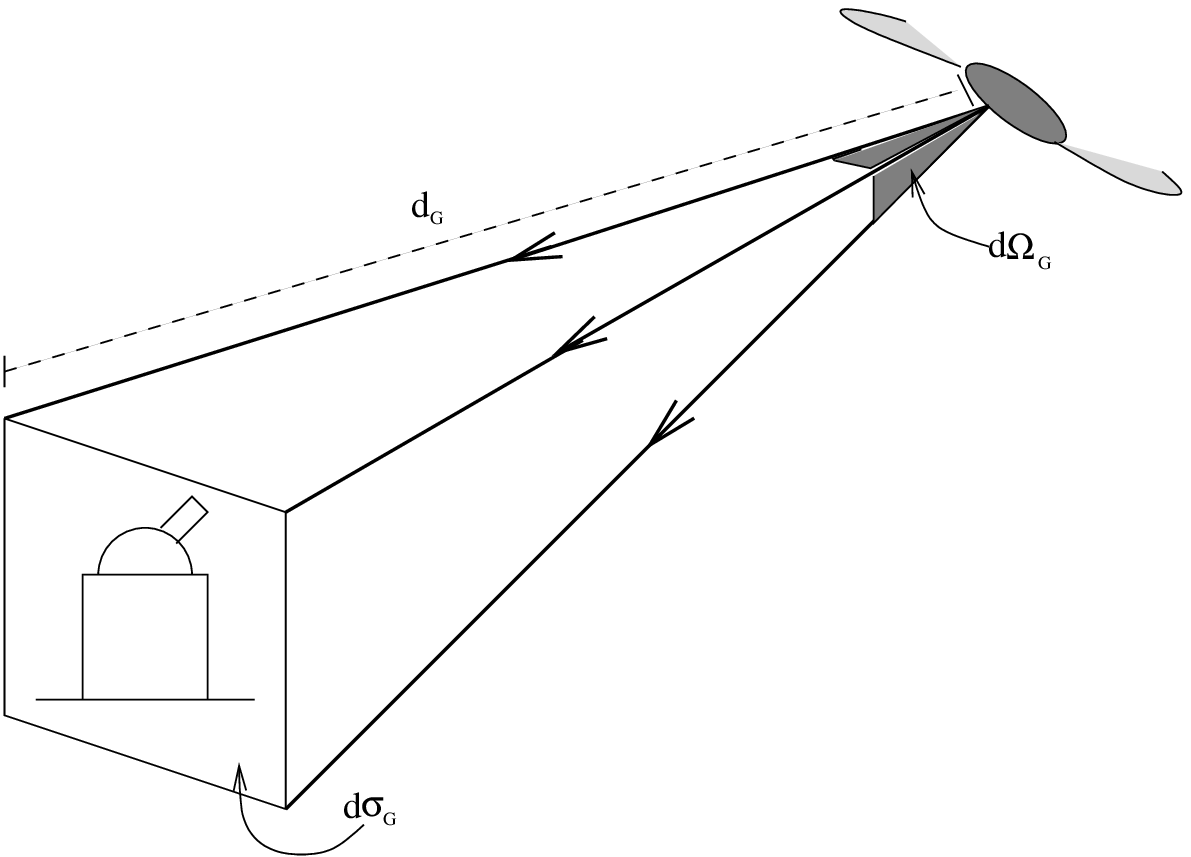}
      \caption{The galaxy area distance $\dg$ is obtained by the
               relationship between the cross-sectional area $\dsg$
	       measured at the observer's rest-frame, and the solid
	       angle $\dog$ measured at the source's rest-frame
	       (eq.\ \ref{dg}). In the Einstein-de Sitter cosmological
	       model this is the same distance as given by Mattig's
	       formula, also having the same behaviour against $z$
	       as the comoving distance (see \S \ref{tdis}).}
      \lb{figdg}
\end{figure*}
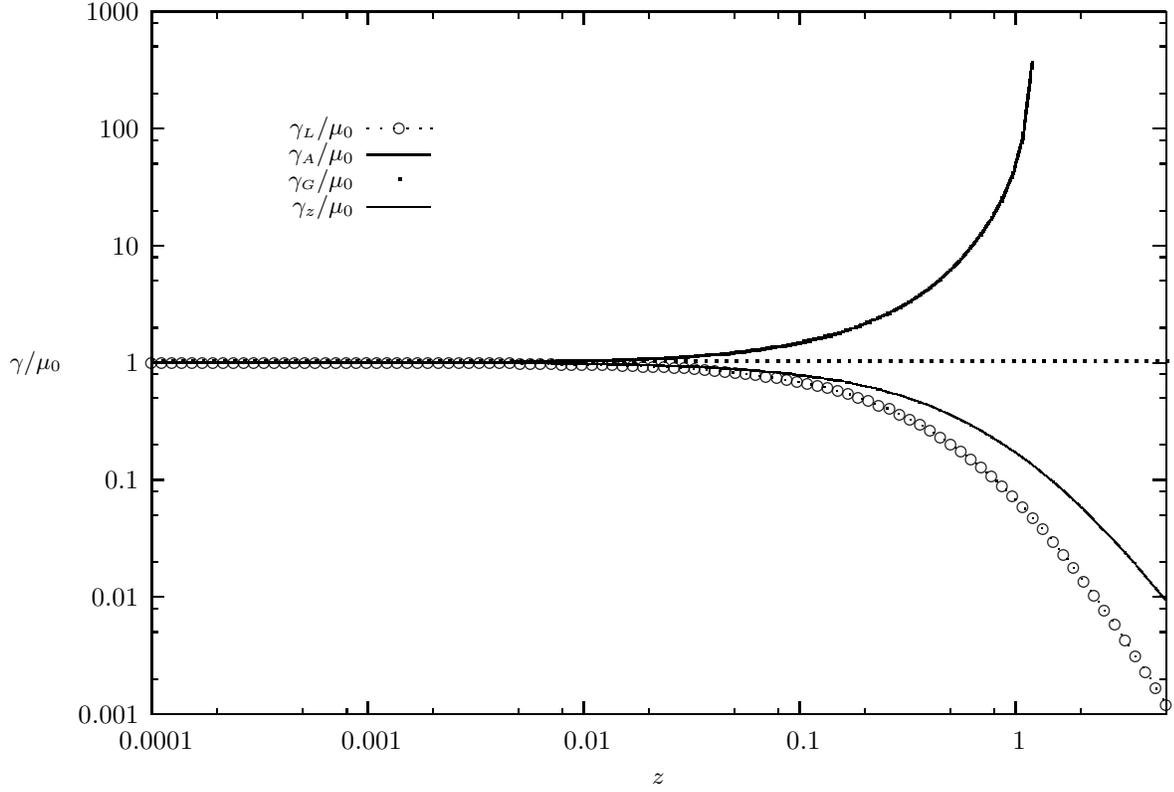
\begin{figure*}
      \centering
      \input{gama.tex}
      \caption{Plot of the differential density $\gamma$ for each
      observational distance in EdS cosmology. Here $\mu_0$ is a
      constant factor related to the Hubble constant and galactic
      rest mass (see eq.\ \ref{mu0}). The proper and comoving
      distances are included, since the former is equal to $\da$,
      whereas the latter has the same behaviour against $z$ as $\dg$.
      Notice that the distance obtained by means of Mattig's
      formula is equivalent to choosing $\dg$ (see \S \ref{tdis}). The
      curves show very clearly the strong dependence of $\gamma$ with
      the chosen distance. $\gamma_{\ssty A}$ diverges at $z=1.25$ and
      loses then its analytical significance for higher $z$. Notice too
      that by mixing up the opposite behaviour of $\gamma$ as given by
      $\da$ and $\dl$, CT01 have in fact smoothed out both the increase
      and decrease of $\gamma_{\ssty A}$ and $\gamma_{\ssty L}$
      respectively, explaining then the odd behaviour obtained by them.}
      \lb{fig-gama}
\end{figure*}
\begin{figure*}
      \centering
      \input{gama-s.tex}
      \caption{Plot of the integral differential density $\gamma^\ast$
               for each observational distance in EdS cosmological model
	       ($\mu_0$ is a constant factor defined in eq.\ \ref{mu0}).
	       Both the comoving and the proper distances are included
	       as they are respectively equivalent to $\dg$ and $\da$.
	       Again, the choice of distance fundamentally changes the
	       behaviour of $\gamma^\ast$ for higher redshifts. The
	       homogeneous situation is obtained when one adopts the
	       galaxy area distance $\dg$, or the comoving distance,
	       or, still, the distance given by Mattig's formula
	       (see \S \ref{tdis}). Notice that a significant
	       deviation from the homogeneous case occurs at the
	       relatively low value of $z=0.03$ ($\approx$ 100Mpc)
	       when $\gamma_{\ssty G}^\ast$ remains constant whereas
	       all others do not. Since the integral differential density
	       is the same as the average number density (see
	       eqs.\ \ref{nl}--\ref{nz}), $\gamma^\ast$ is the
	       best statistical test for general tendencies of the
	       galaxy clumping behaviour.}
      \lb{fig-gama-s}
\end{figure*}
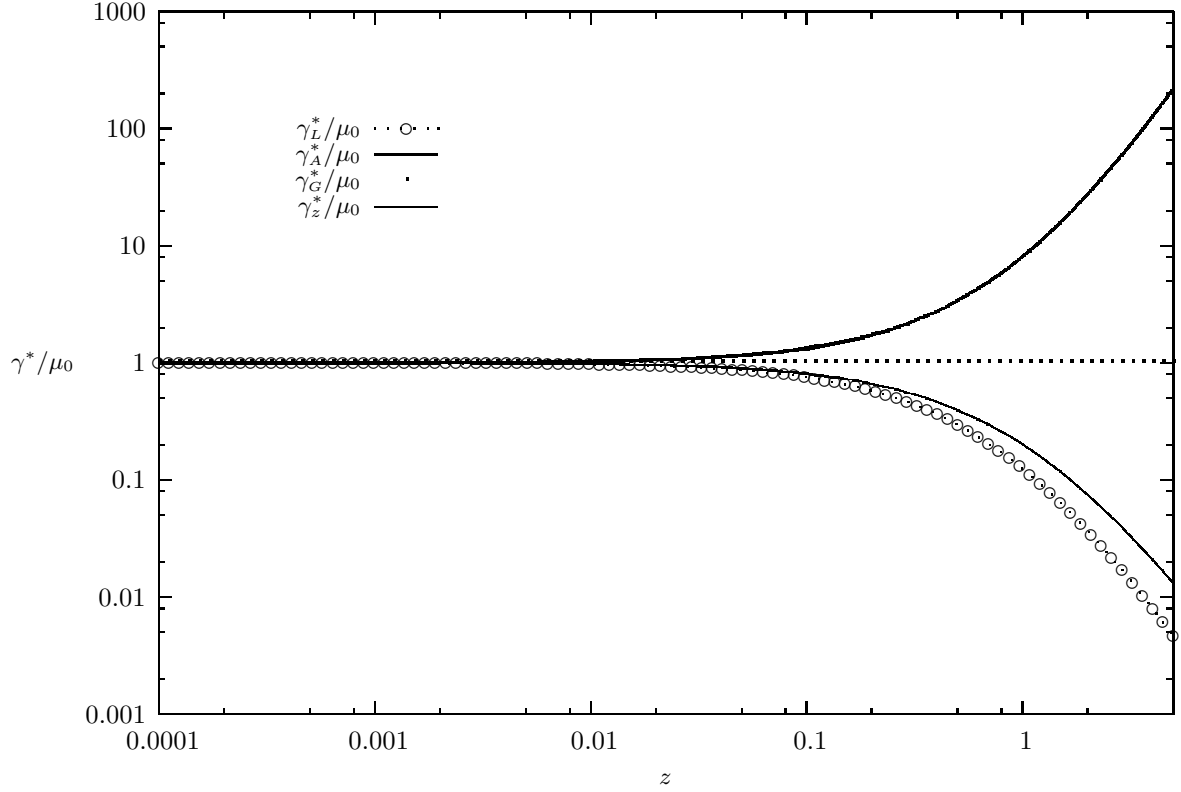

\end{document}

%% file: gama.tex
\setlength{\unitlength}{0.240900pt}
\ifx\plotpoint\undefined\newsavebox{\plotpoint}\fi
\sbox{\plotpoint}{\rule[-0.200pt]{0.400pt}{0.400pt}}%
\begin{picture}(1875,1350)(0,0)
\font\gnuplot=cmr10 at 10pt
\gnuplot
\sbox{\plotpoint}{\rule[-0.200pt]{0.400pt}{0.400pt}}%
\put(221.0,123.0){\rule[-0.200pt]{4.818pt}{0.400pt}}
\put(201,123){\makebox(0,0)[r]{ 0.001}}
\put(1794.0,123.0){\rule[-0.200pt]{4.818pt}{0.400pt}}
\put(221.0,178.0){\rule[-0.200pt]{2.409pt}{0.400pt}}
\put(1804.0,178.0){\rule[-0.200pt]{2.409pt}{0.400pt}}
\put(221.0,252.0){\rule[-0.200pt]{2.409pt}{0.400pt}}
\put(1804.0,252.0){\rule[-0.200pt]{2.409pt}{0.400pt}}
\put(221.0,289.0){\rule[-0.200pt]{2.409pt}{0.400pt}}
\put(1804.0,289.0){\rule[-0.200pt]{2.409pt}{0.400pt}}
\put(221.0,307.0){\rule[-0.200pt]{4.818pt}{0.400pt}}
\put(201,307){\makebox(0,0)[r]{ 0.01}}
\put(1794.0,307.0){\rule[-0.200pt]{4.818pt}{0.400pt}}
\put(221.0,362.0){\rule[-0.200pt]{2.409pt}{0.400pt}}
\put(1804.0,362.0){\rule[-0.200pt]{2.409pt}{0.400pt}}
\put(221.0,436.0){\rule[-0.200pt]{2.409pt}{0.400pt}}
\put(1804.0,436.0){\rule[-0.200pt]{2.409pt}{0.400pt}}
\put(221.0,473.0){\rule[-0.200pt]{2.409pt}{0.400pt}}
\put(1804.0,473.0){\rule[-0.200pt]{2.409pt}{0.400pt}}
\put(221.0,491.0){\rule[-0.200pt]{4.818pt}{0.400pt}}
\put(201,491){\makebox(0,0)[r]{ 0.1}}
\put(1794.0,491.0){\rule[-0.200pt]{4.818pt}{0.400pt}}
\put(221.0,546.0){\rule[-0.200pt]{2.409pt}{0.400pt}}
\put(1804.0,546.0){\rule[-0.200pt]{2.409pt}{0.400pt}}
\put(221.0,620.0){\rule[-0.200pt]{2.409pt}{0.400pt}}
\put(1804.0,620.0){\rule[-0.200pt]{2.409pt}{0.400pt}}
\put(221.0,657.0){\rule[-0.200pt]{2.409pt}{0.400pt}}
\put(1804.0,657.0){\rule[-0.200pt]{2.409pt}{0.400pt}}
\put(221.0,675.0){\rule[-0.200pt]{4.818pt}{0.400pt}}
\put(201,675){\makebox(0,0)[r]{ 1}}
\put(1794.0,675.0){\rule[-0.200pt]{4.818pt}{0.400pt}}
\put(221.0,730.0){\rule[-0.200pt]{2.409pt}{0.400pt}}
\put(1804.0,730.0){\rule[-0.200pt]{2.409pt}{0.400pt}}
\put(221.0,804.0){\rule[-0.200pt]{2.409pt}{0.400pt}}
\put(1804.0,804.0){\rule[-0.200pt]{2.409pt}{0.400pt}}
\put(221.0,841.0){\rule[-0.200pt]{2.409pt}{0.400pt}}
\put(1804.0,841.0){\rule[-0.200pt]{2.409pt}{0.400pt}}
\put(221.0,859.0){\rule[-0.200pt]{4.818pt}{0.400pt}}
\put(201,859){\makebox(0,0)[r]{ 10}}
\put(1794.0,859.0){\rule[-0.200pt]{4.818pt}{0.400pt}}
\put(221.0,914.0){\rule[-0.200pt]{2.409pt}{0.400pt}}
\put(1804.0,914.0){\rule[-0.200pt]{2.409pt}{0.400pt}}
\put(221.0,988.0){\rule[-0.200pt]{2.409pt}{0.400pt}}
\put(1804.0,988.0){\rule[-0.200pt]{2.409pt}{0.400pt}}
\put(221.0,1025.0){\rule[-0.200pt]{2.409pt}{0.400pt}}
\put(1804.0,1025.0){\rule[-0.200pt]{2.409pt}{0.400pt}}
\put(221.0,1043.0){\rule[-0.200pt]{4.818pt}{0.400pt}}
\put(201,1043){\makebox(0,0)[r]{ 100}}
\put(1794.0,1043.0){\rule[-0.200pt]{4.818pt}{0.400pt}}
\put(221.0,1098.0){\rule[-0.200pt]{2.409pt}{0.400pt}}
\put(1804.0,1098.0){\rule[-0.200pt]{2.409pt}{0.400pt}}
\put(221.0,1172.0){\rule[-0.200pt]{2.409pt}{0.400pt}}
\put(1804.0,1172.0){\rule[-0.200pt]{2.409pt}{0.400pt}}
\put(221.0,1209.0){\rule[-0.200pt]{2.409pt}{0.400pt}}
\put(1804.0,1209.0){\rule[-0.200pt]{2.409pt}{0.400pt}}
\put(221.0,1227.0){\rule[-0.200pt]{4.818pt}{0.400pt}}
\put(201,1227){\makebox(0,0)[r]{ 1000}}
\put(1794.0,1227.0){\rule[-0.200pt]{4.818pt}{0.400pt}}
\put(221.0,123.0){\rule[-0.200pt]{0.400pt}{4.818pt}}
\put(221,82){\makebox(0,0){ 0.0001}}
\put(221.0,1207.0){\rule[-0.200pt]{0.400pt}{4.818pt}}
\put(323.0,123.0){\rule[-0.200pt]{0.400pt}{2.409pt}}
\put(323.0,1217.0){\rule[-0.200pt]{0.400pt}{2.409pt}}
\put(458.0,123.0){\rule[-0.200pt]{0.400pt}{2.409pt}}
\put(458.0,1217.0){\rule[-0.200pt]{0.400pt}{2.409pt}}
\put(527.0,123.0){\rule[-0.200pt]{0.400pt}{2.409pt}}
\put(527.0,1217.0){\rule[-0.200pt]{0.400pt}{2.409pt}}
\put(560.0,123.0){\rule[-0.200pt]{0.400pt}{4.818pt}}
\put(560,82){\makebox(0,0){ 0.001}}
\put(560.0,1207.0){\rule[-0.200pt]{0.400pt}{4.818pt}}
\put(662.0,123.0){\rule[-0.200pt]{0.400pt}{2.409pt}}
\put(662.0,1217.0){\rule[-0.200pt]{0.400pt}{2.409pt}}
\put(797.0,123.0){\rule[-0.200pt]{0.400pt}{2.409pt}}
\put(797.0,1217.0){\rule[-0.200pt]{0.400pt}{2.409pt}}
\put(866.0,123.0){\rule[-0.200pt]{0.400pt}{2.409pt}}
\put(866.0,1217.0){\rule[-0.200pt]{0.400pt}{2.409pt}}
\put(899.0,123.0){\rule[-0.200pt]{0.400pt}{4.818pt}}
\put(899,82){\makebox(0,0){ 0.01}}
\put(899.0,1207.0){\rule[-0.200pt]{0.400pt}{4.818pt}}
\put(1001.0,123.0){\rule[-0.200pt]{0.400pt}{2.409pt}}
\put(1001.0,1217.0){\rule[-0.200pt]{0.400pt}{2.409pt}}
\put(1136.0,123.0){\rule[-0.200pt]{0.400pt}{2.409pt}}
\put(1136.0,1217.0){\rule[-0.200pt]{0.400pt}{2.409pt}}
\put(1205.0,123.0){\rule[-0.200pt]{0.400pt}{2.409pt}}
\put(1205.0,1217.0){\rule[-0.200pt]{0.400pt}{2.409pt}}
\put(1238.0,123.0){\rule[-0.200pt]{0.400pt}{4.818pt}}
\put(1238,82){\makebox(0,0){ 0.1}}
\put(1238.0,1207.0){\rule[-0.200pt]{0.400pt}{4.818pt}}
\put(1340.0,123.0){\rule[-0.200pt]{0.400pt}{2.409pt}}
\put(1340.0,1217.0){\rule[-0.200pt]{0.400pt}{2.409pt}}
\put(1475.0,123.0){\rule[-0.200pt]{0.400pt}{2.409pt}}
\put(1475.0,1217.0){\rule[-0.200pt]{0.400pt}{2.409pt}}
\put(1544.0,123.0){\rule[-0.200pt]{0.400pt}{2.409pt}}
\put(1544.0,1217.0){\rule[-0.200pt]{0.400pt}{2.409pt}}
\put(1577.0,123.0){\rule[-0.200pt]{0.400pt}{4.818pt}}
\put(1577,82){\makebox(0,0){ 1}}
\put(1577.0,1207.0){\rule[-0.200pt]{0.400pt}{4.818pt}}
\put(1679.0,123.0){\rule[-0.200pt]{0.400pt}{2.409pt}}
\put(1679.0,1217.0){\rule[-0.200pt]{0.400pt}{2.409pt}}
\put(1814.0,123.0){\rule[-0.200pt]{0.400pt}{2.409pt}}
\put(1814.0,1217.0){\rule[-0.200pt]{0.400pt}{2.409pt}}
\put(221.0,123.0){\rule[-0.200pt]{383.754pt}{0.400pt}}
\put(1814.0,123.0){\rule[-0.200pt]{0.400pt}{265.954pt}}
\put(221.0,1227.0){\rule[-0.200pt]{383.754pt}{0.400pt}}
\put(40,675){\makebox(0,0){$\gamma / \mu_0 $}}
\put(1017,21){\makebox(0,0){$z$}}
\put(1017,1289){\makebox(0,0){ }}
\put(221.0,123.0){\rule[-0.200pt]{0.400pt}{265.954pt}}
\put(540,1043){\makebox(0,0)[r]{$\gamma_{\ssty L}/\mu_0$}}
\multiput(560,1043)(20.756,0.000){5}{\usebox{\plotpoint}}
\put(660,1043){\usebox{\plotpoint}}
\put(221,675){\usebox{\plotpoint}}
\put(221.00,675.00){\usebox{\plotpoint}}
\put(241.76,675.00){\usebox{\plotpoint}}
\put(262.51,675.00){\usebox{\plotpoint}}
\put(283.27,675.00){\usebox{\plotpoint}}
\put(304.02,675.00){\usebox{\plotpoint}}
\put(324.78,675.00){\usebox{\plotpoint}}
\put(345.53,675.00){\usebox{\plotpoint}}
\put(366.29,675.00){\usebox{\plotpoint}}
\put(387.04,675.00){\usebox{\plotpoint}}
\put(407.80,675.00){\usebox{\plotpoint}}
\put(428.55,675.00){\usebox{\plotpoint}}
\put(449.31,675.00){\usebox{\plotpoint}}
\put(470.07,675.00){\usebox{\plotpoint}}
\put(490.82,675.00){\usebox{\plotpoint}}
\put(511.58,675.00){\usebox{\plotpoint}}
\put(532.33,675.00){\usebox{\plotpoint}}
\put(553.09,675.00){\usebox{\plotpoint}}
\put(573.84,675.00){\usebox{\plotpoint}}
\put(594.60,675.00){\usebox{\plotpoint}}
\put(615.35,675.00){\usebox{\plotpoint}}
\put(636.08,674.18){\usebox{\plotpoint}}
\put(656.83,674.00){\usebox{\plotpoint}}
\put(677.59,674.00){\usebox{\plotpoint}}
\put(698.35,674.00){\usebox{\plotpoint}}
\put(719.10,674.00){\usebox{\plotpoint}}
\put(739.86,674.00){\usebox{\plotpoint}}
\put(760.61,674.00){\usebox{\plotpoint}}
\put(781.37,674.00){\usebox{\plotpoint}}
\put(802.09,673.00){\usebox{\plotpoint}}
\put(822.85,673.00){\usebox{\plotpoint}}
\put(843.60,673.00){\usebox{\plotpoint}}
\put(864.33,672.04){\usebox{\plotpoint}}
\put(885.08,672.00){\usebox{\plotpoint}}
\put(905.84,672.00){\usebox{\plotpoint}}
\put(926.57,671.15){\usebox{\plotpoint}}
\put(947.31,670.86){\usebox{\plotpoint}}
\put(968.04,670.00){\usebox{\plotpoint}}
\put(988.77,669.26){\usebox{\plotpoint}}
\put(1009.49,668.00){\usebox{\plotpoint}}
\put(1030.24,667.74){\usebox{\plotpoint}}
\put(1050.95,666.44){\usebox{\plotpoint}}
\put(1071.67,665.15){\usebox{\plotpoint}}
\put(1092.37,663.70){\usebox{\plotpoint}}
\put(1113.00,661.56){\usebox{\plotpoint}}
\put(1133.65,659.54){\usebox{\plotpoint}}
\put(1154.25,656.97){\usebox{\plotpoint}}
\put(1174.84,654.39){\usebox{\plotpoint}}
\put(1195.45,651.89){\usebox{\plotpoint}}
\put(1215.92,648.58){\usebox{\plotpoint}}
\put(1236.32,644.75){\usebox{\plotpoint}}
\put(1256.72,640.93){\usebox{\plotpoint}}
\put(1276.99,636.50){\usebox{\plotpoint}}
\put(1296.90,630.66){\usebox{\plotpoint}}
\put(1316.93,625.28){\usebox{\plotpoint}}
\put(1336.47,618.29){\usebox{\plotpoint}}
\put(1355.91,611.10){\usebox{\plotpoint}}
\put(1375.27,603.67){\usebox{\plotpoint}}
\put(1394.08,594.96){\usebox{\plotpoint}}
\put(1412.65,585.68){\usebox{\plotpoint}}
\put(1431.04,576.10){\usebox{\plotpoint}}
\put(1448.64,565.10){\usebox{\plotpoint}}
\put(1466.07,553.83){\usebox{\plotpoint}}
\put(1483.17,542.07){\usebox{\plotpoint}}
\put(1499.79,529.67){\usebox{\plotpoint}}
\put(1515.90,516.58){\usebox{\plotpoint}}
\put(1531.77,503.20){\usebox{\plotpoint}}
\put(1547.36,489.51){\usebox{\plotpoint}}
\put(1562.59,475.41){\usebox{\plotpoint}}
\put(1577.13,460.61){\usebox{\plotpoint}}
\put(1591.36,445.50){\usebox{\plotpoint}}
\multiput(1605,431)(13.789,-15.513){2}{\usebox{\plotpoint}}
\put(1632.42,398.73){\usebox{\plotpoint}}
\put(1645.38,382.52){\usebox{\plotpoint}}
\put(1658.35,366.31){\usebox{\plotpoint}}
\multiput(1669,353)(12.208,-16.786){2}{\usebox{\plotpoint}}
\put(1695.60,316.43){\usebox{\plotpoint}}
\put(1707.61,299.51){\usebox{\plotpoint}}
\multiput(1717,286)(12.337,-16.691){2}{\usebox{\plotpoint}}
\put(1743.55,248.68){\usebox{\plotpoint}}
\put(1754.92,231.31){\usebox{\plotpoint}}
\multiput(1766,214)(11.188,-17.482){2}{\usebox{\plotpoint}}
\put(1788.30,178.76){\usebox{\plotpoint}}
\multiput(1798,163)(10.878,-17.677){2}{\usebox{\plotpoint}}
\put(1814,137){\usebox{\plotpoint}}
\put(221,675){\circle{18}}
\put(237,675){\circle{18}}
\put(253,675){\circle{18}}
\put(269,675){\circle{18}}
\put(285,675){\circle{18}}
\put(301,675){\circle{18}}
\put(318,675){\circle{18}}
\put(334,675){\circle{18}}
\put(350,675){\circle{18}}
\put(366,675){\circle{18}}
\put(382,675){\circle{18}}
\put(398,675){\circle{18}}
\put(414,675){\circle{18}}
\put(430,675){\circle{18}}
\put(446,675){\circle{18}}
\put(462,675){\circle{18}}
\put(478,675){\circle{18}}
\put(495,675){\circle{18}}
\put(511,675){\circle{18}}
\put(527,675){\circle{18}}
\put(543,675){\circle{18}}
\put(559,675){\circle{18}}
\put(575,675){\circle{18}}
\put(591,675){\circle{18}}
\put(607,675){\circle{18}}
\put(623,675){\circle{18}}
\put(639,674){\circle{18}}
\put(655,674){\circle{18}}
\put(672,674){\circle{18}}
\put(688,674){\circle{18}}
\put(704,674){\circle{18}}
\put(720,674){\circle{18}}
\put(736,674){\circle{18}}
\put(752,674){\circle{18}}
\put(768,674){\circle{18}}
\put(784,674){\circle{18}}
\put(800,673){\circle{18}}
\put(816,673){\circle{18}}
\put(832,673){\circle{18}}
\put(849,673){\circle{18}}
\put(865,672){\circle{18}}
\put(881,672){\circle{18}}
\put(897,672){\circle{18}}
\put(913,672){\circle{18}}
\put(929,671){\circle{18}}
\put(945,671){\circle{18}}
\put(961,670){\circle{18}}
\put(977,670){\circle{18}}
\put(993,669){\circle{18}}
\put(1009,668){\circle{18}}
\put(1026,668){\circle{18}}
\put(1042,667){\circle{18}}
\put(1058,666){\circle{18}}
\put(1074,665){\circle{18}}
\put(1090,664){\circle{18}}
\put(1106,662){\circle{18}}
\put(1122,661){\circle{18}}
\put(1138,659){\circle{18}}
\put(1154,657){\circle{18}}
\put(1170,655){\circle{18}}
\put(1186,653){\circle{18}}
\put(1203,651){\circle{18}}
\put(1219,648){\circle{18}}
\put(1235,645){\circle{18}}
\put(1251,642){\circle{18}}
\put(1267,639){\circle{18}}
\put(1283,635){\circle{18}}
\put(1299,630){\circle{18}}
\put(1315,626){\circle{18}}
\put(1331,620){\circle{18}}
\put(1347,615){\circle{18}}
\put(1363,608){\circle{18}}
\put(1380,602){\circle{18}}
\put(1396,594){\circle{18}}
\put(1412,586){\circle{18}}
\put(1428,578){\circle{18}}
\put(1444,568){\circle{18}}
\put(1460,558){\circle{18}}
\put(1476,547){\circle{18}}
\put(1492,536){\circle{18}}
\put(1508,523){\circle{18}}
\put(1524,510){\circle{18}}
\put(1540,496){\circle{18}}
\put(1557,481){\circle{18}}
\put(1573,465){\circle{18}}
\put(1589,448){\circle{18}}
\put(1605,431){\circle{18}}
\put(1621,413){\circle{18}}
\put(1637,393){\circle{18}}
\put(1653,373){\circle{18}}
\put(1669,353){\circle{18}}
\put(1685,331){\circle{18}}
\put(1701,309){\circle{18}}
\put(1717,286){\circle{18}}
\put(1734,263){\circle{18}}
\put(1750,239){\circle{18}}
\put(1766,214){\circle{18}}
\put(1782,189){\circle{18}}
\put(1798,163){\circle{18}}
\put(1814,137){\circle{18}}
\put(610,1043){\circle{18}}
\sbox{\plotpoint}{\rule[-0.400pt]{0.800pt}{0.800pt}}%
\put(540,1002){\makebox(0,0)[r]{$\gamma_{\ssty A}/\mu_0$}}
\put(560.0,1002.0){\rule[-0.400pt]{24.090pt}{0.800pt}}
\put(221,675){\usebox{\plotpoint}}
\put(623,673.84){\rule{3.854pt}{0.800pt}}
\multiput(623.00,673.34)(8.000,1.000){2}{\rule{1.927pt}{0.800pt}}
\put(221.0,675.0){\rule[-0.400pt]{96.842pt}{0.800pt}}
\put(784,674.84){\rule{3.854pt}{0.800pt}}
\multiput(784.00,674.34)(8.000,1.000){2}{\rule{1.927pt}{0.800pt}}
\put(639.0,676.0){\rule[-0.400pt]{34.930pt}{0.800pt}}
\put(849,675.84){\rule{3.854pt}{0.800pt}}
\multiput(849.00,675.34)(8.000,1.000){2}{\rule{1.927pt}{0.800pt}}
\put(800.0,677.0){\rule[-0.400pt]{11.804pt}{0.800pt}}
\put(897,676.84){\rule{3.854pt}{0.800pt}}
\multiput(897.00,676.34)(8.000,1.000){2}{\rule{1.927pt}{0.800pt}}
\put(865.0,678.0){\rule[-0.400pt]{7.709pt}{0.800pt}}
\put(945,677.84){\rule{3.854pt}{0.800pt}}
\multiput(945.00,677.34)(8.000,1.000){2}{\rule{1.927pt}{0.800pt}}
\put(913.0,679.0){\rule[-0.400pt]{7.709pt}{0.800pt}}
\put(977,678.84){\rule{3.854pt}{0.800pt}}
\multiput(977.00,678.34)(8.000,1.000){2}{\rule{1.927pt}{0.800pt}}
\put(993,679.84){\rule{3.854pt}{0.800pt}}
\multiput(993.00,679.34)(8.000,1.000){2}{\rule{1.927pt}{0.800pt}}
\put(961.0,680.0){\rule[-0.400pt]{3.854pt}{0.800pt}}
\put(1026,680.84){\rule{3.854pt}{0.800pt}}
\multiput(1026.00,680.34)(8.000,1.000){2}{\rule{1.927pt}{0.800pt}}
\put(1042,681.84){\rule{3.854pt}{0.800pt}}
\multiput(1042.00,681.34)(8.000,1.000){2}{\rule{1.927pt}{0.800pt}}
\put(1058,682.84){\rule{3.854pt}{0.800pt}}
\multiput(1058.00,682.34)(8.000,1.000){2}{\rule{1.927pt}{0.800pt}}
\put(1074,684.34){\rule{3.854pt}{0.800pt}}
\multiput(1074.00,683.34)(8.000,2.000){2}{\rule{1.927pt}{0.800pt}}
\put(1090,685.84){\rule{3.854pt}{0.800pt}}
\multiput(1090.00,685.34)(8.000,1.000){2}{\rule{1.927pt}{0.800pt}}
\put(1106,686.84){\rule{3.854pt}{0.800pt}}
\multiput(1106.00,686.34)(8.000,1.000){2}{\rule{1.927pt}{0.800pt}}
\put(1122,688.34){\rule{3.854pt}{0.800pt}}
\multiput(1122.00,687.34)(8.000,2.000){2}{\rule{1.927pt}{0.800pt}}
\put(1138,690.34){\rule{3.854pt}{0.800pt}}
\multiput(1138.00,689.34)(8.000,2.000){2}{\rule{1.927pt}{0.800pt}}
\put(1154,692.34){\rule{3.854pt}{0.800pt}}
\multiput(1154.00,691.34)(8.000,2.000){2}{\rule{1.927pt}{0.800pt}}
\put(1170,694.34){\rule{3.854pt}{0.800pt}}
\multiput(1170.00,693.34)(8.000,2.000){2}{\rule{1.927pt}{0.800pt}}
\put(1186,696.84){\rule{4.095pt}{0.800pt}}
\multiput(1186.00,695.34)(8.500,3.000){2}{\rule{2.048pt}{0.800pt}}
\put(1203,699.34){\rule{3.854pt}{0.800pt}}
\multiput(1203.00,698.34)(8.000,2.000){2}{\rule{1.927pt}{0.800pt}}
\put(1219,701.84){\rule{3.854pt}{0.800pt}}
\multiput(1219.00,700.34)(8.000,3.000){2}{\rule{1.927pt}{0.800pt}}
\put(1235,705.34){\rule{3.400pt}{0.800pt}}
\multiput(1235.00,703.34)(8.943,4.000){2}{\rule{1.700pt}{0.800pt}}
\put(1251,709.34){\rule{3.400pt}{0.800pt}}
\multiput(1251.00,707.34)(8.943,4.000){2}{\rule{1.700pt}{0.800pt}}
\put(1267,713.34){\rule{3.400pt}{0.800pt}}
\multiput(1267.00,711.34)(8.943,4.000){2}{\rule{1.700pt}{0.800pt}}
\put(1283,717.34){\rule{3.400pt}{0.800pt}}
\multiput(1283.00,715.34)(8.943,4.000){2}{\rule{1.700pt}{0.800pt}}
\multiput(1299.00,722.39)(1.579,0.536){5}{\rule{2.333pt}{0.129pt}}
\multiput(1299.00,719.34)(11.157,6.000){2}{\rule{1.167pt}{0.800pt}}
\multiput(1315.00,728.38)(2.271,0.560){3}{\rule{2.760pt}{0.135pt}}
\multiput(1315.00,725.34)(10.271,5.000){2}{\rule{1.380pt}{0.800pt}}
\multiput(1331.00,733.40)(1.263,0.526){7}{\rule{2.029pt}{0.127pt}}
\multiput(1331.00,730.34)(11.790,7.000){2}{\rule{1.014pt}{0.800pt}}
\multiput(1347.00,740.40)(1.263,0.526){7}{\rule{2.029pt}{0.127pt}}
\multiput(1347.00,737.34)(11.790,7.000){2}{\rule{1.014pt}{0.800pt}}
\multiput(1363.00,747.40)(1.351,0.526){7}{\rule{2.143pt}{0.127pt}}
\multiput(1363.00,744.34)(12.552,7.000){2}{\rule{1.071pt}{0.800pt}}
\multiput(1380.00,754.40)(0.927,0.516){11}{\rule{1.622pt}{0.124pt}}
\multiput(1380.00,751.34)(12.633,9.000){2}{\rule{0.811pt}{0.800pt}}
\multiput(1396.00,763.40)(0.927,0.516){11}{\rule{1.622pt}{0.124pt}}
\multiput(1396.00,760.34)(12.633,9.000){2}{\rule{0.811pt}{0.800pt}}
\multiput(1412.00,772.40)(0.739,0.512){15}{\rule{1.364pt}{0.123pt}}
\multiput(1412.00,769.34)(13.170,11.000){2}{\rule{0.682pt}{0.800pt}}
\multiput(1428.00,783.40)(0.739,0.512){15}{\rule{1.364pt}{0.123pt}}
\multiput(1428.00,780.34)(13.170,11.000){2}{\rule{0.682pt}{0.800pt}}
\multiput(1444.00,794.41)(0.616,0.509){19}{\rule{1.185pt}{0.123pt}}
\multiput(1444.00,791.34)(13.541,13.000){2}{\rule{0.592pt}{0.800pt}}
\multiput(1460.00,807.41)(0.529,0.508){23}{\rule{1.053pt}{0.122pt}}
\multiput(1460.00,804.34)(13.814,15.000){2}{\rule{0.527pt}{0.800pt}}
\multiput(1476.00,822.41)(0.494,0.507){25}{\rule{1.000pt}{0.122pt}}
\multiput(1476.00,819.34)(13.924,16.000){2}{\rule{0.500pt}{0.800pt}}
\multiput(1493.41,837.00)(0.507,0.593){25}{\rule{0.122pt}{1.150pt}}
\multiput(1490.34,837.00)(16.000,16.613){2}{\rule{0.800pt}{0.575pt}}
\multiput(1509.41,856.00)(0.507,0.659){25}{\rule{0.122pt}{1.250pt}}
\multiput(1506.34,856.00)(16.000,18.406){2}{\rule{0.800pt}{0.625pt}}
\multiput(1525.41,877.00)(0.507,0.758){25}{\rule{0.122pt}{1.400pt}}
\multiput(1522.34,877.00)(16.000,21.094){2}{\rule{0.800pt}{0.700pt}}
\multiput(1541.41,901.00)(0.507,0.897){27}{\rule{0.122pt}{1.612pt}}
\multiput(1538.34,901.00)(17.000,26.655){2}{\rule{0.800pt}{0.806pt}}
\multiput(1558.41,931.00)(0.507,1.220){25}{\rule{0.122pt}{2.100pt}}
\multiput(1555.34,931.00)(16.000,33.641){2}{\rule{0.800pt}{1.050pt}}
\multiput(1574.41,969.00)(0.507,1.814){25}{\rule{0.122pt}{3.000pt}}
\multiput(1571.34,969.00)(16.000,49.773){2}{\rule{0.800pt}{1.500pt}}
\multiput(1590.41,1025.00)(0.507,4.060){25}{\rule{0.122pt}{6.400pt}}
\multiput(1587.34,1025.00)(16.000,110.716){2}{\rule{0.800pt}{3.200pt}}
\put(1009.0,682.0){\rule[-0.400pt]{4.095pt}{0.800pt}}
\sbox{\plotpoint}{\rule[-0.200pt]{0.400pt}{0.400pt}}%
\put(540,961){\makebox(0,0)[r]{$\gamma_{\ssty G}/\mu_0$}}
\put(610,961){\rule{1pt}{1pt}}
\put(221,675){\rule{1pt}{1pt}}
\put(237,675){\rule{1pt}{1pt}}
\put(253,675){\rule{1pt}{1pt}}
\put(269,675){\rule{1pt}{1pt}}
\put(285,675){\rule{1pt}{1pt}}
\put(301,675){\rule{1pt}{1pt}}
\put(318,675){\rule{1pt}{1pt}}
\put(334,675){\rule{1pt}{1pt}}
\put(350,675){\rule{1pt}{1pt}}
\put(366,675){\rule{1pt}{1pt}}
\put(382,675){\rule{1pt}{1pt}}
\put(398,675){\rule{1pt}{1pt}}
\put(414,675){\rule{1pt}{1pt}}
\put(430,675){\rule{1pt}{1pt}}
\put(446,675){\rule{1pt}{1pt}}
\put(462,675){\rule{1pt}{1pt}}
\put(478,675){\rule{1pt}{1pt}}
\put(495,675){\rule{1pt}{1pt}}
\put(511,675){\rule{1pt}{1pt}}
\put(527,675){\rule{1pt}{1pt}}
\put(543,675){\rule{1pt}{1pt}}
\put(559,675){\rule{1pt}{1pt}}
\put(575,675){\rule{1pt}{1pt}}
\put(591,675){\rule{1pt}{1pt}}
\put(607,675){\rule{1pt}{1pt}}
\put(623,675){\rule{1pt}{1pt}}
\put(639,675){\rule{1pt}{1pt}}
\put(655,675){\rule{1pt}{1pt}}
\put(672,675){\rule{1pt}{1pt}}
\put(688,675){\rule{1pt}{1pt}}
\put(704,675){\rule{1pt}{1pt}}
\put(720,675){\rule{1pt}{1pt}}
\put(736,675){\rule{1pt}{1pt}}
\put(752,675){\rule{1pt}{1pt}}
\put(768,675){\rule{1pt}{1pt}}
\put(784,675){\rule{1pt}{1pt}}
\put(800,675){\rule{1pt}{1pt}}
\put(816,675){\rule{1pt}{1pt}}
\put(832,675){\rule{1pt}{1pt}}
\put(849,675){\rule{1pt}{1pt}}
\put(865,675){\rule{1pt}{1pt}}
\put(881,675){\rule{1pt}{1pt}}
\put(897,675){\rule{1pt}{1pt}}
\put(913,675){\rule{1pt}{1pt}}
\put(929,675){\rule{1pt}{1pt}}
\put(945,675){\rule{1pt}{1pt}}
\put(961,675){\rule{1pt}{1pt}}
\put(977,675){\rule{1pt}{1pt}}
\put(993,675){\rule{1pt}{1pt}}
\put(1009,675){\rule{1pt}{1pt}}
\put(1026,675){\rule{1pt}{1pt}}
\put(1042,675){\rule{1pt}{1pt}}
\put(1058,675){\rule{1pt}{1pt}}
\put(1074,675){\rule{1pt}{1pt}}
\put(1090,675){\rule{1pt}{1pt}}
\put(1106,675){\rule{1pt}{1pt}}
\put(1122,675){\rule{1pt}{1pt}}
\put(1138,675){\rule{1pt}{1pt}}
\put(1154,675){\rule{1pt}{1pt}}
\put(1170,675){\rule{1pt}{1pt}}
\put(1186,675){\rule{1pt}{1pt}}
\put(1203,675){\rule{1pt}{1pt}}
\put(1219,675){\rule{1pt}{1pt}}
\put(1235,675){\rule{1pt}{1pt}}
\put(1251,675){\rule{1pt}{1pt}}
\put(1267,675){\rule{1pt}{1pt}}
\put(1283,675){\rule{1pt}{1pt}}
\put(1299,675){\rule{1pt}{1pt}}
\put(1315,675){\rule{1pt}{1pt}}
\put(1331,675){\rule{1pt}{1pt}}
\put(1347,675){\rule{1pt}{1pt}}
\put(1363,675){\rule{1pt}{1pt}}
\put(1380,675){\rule{1pt}{1pt}}
\put(1396,675){\rule{1pt}{1pt}}
\put(1412,675){\rule{1pt}{1pt}}
\put(1428,675){\rule{1pt}{1pt}}
\put(1444,675){\rule{1pt}{1pt}}
\put(1460,675){\rule{1pt}{1pt}}
\put(1476,675){\rule{1pt}{1pt}}
\put(1492,675){\rule{1pt}{1pt}}
\put(1508,675){\rule{1pt}{1pt}}
\put(1524,675){\rule{1pt}{1pt}}
\put(1540,675){\rule{1pt}{1pt}}
\put(1557,675){\rule{1pt}{1pt}}
\put(1573,675){\rule{1pt}{1pt}}
\put(1589,675){\rule{1pt}{1pt}}
\put(1605,675){\rule{1pt}{1pt}}
\put(1621,675){\rule{1pt}{1pt}}
\put(1637,675){\rule{1pt}{1pt}}
\put(1653,675){\rule{1pt}{1pt}}
\put(1669,675){\rule{1pt}{1pt}}
\put(1685,675){\rule{1pt}{1pt}}
\put(1701,675){\rule{1pt}{1pt}}
\put(1717,675){\rule{1pt}{1pt}}
\put(1734,675){\rule{1pt}{1pt}}
\put(1750,675){\rule{1pt}{1pt}}
\put(1766,675){\rule{1pt}{1pt}}
\put(1782,675){\rule{1pt}{1pt}}
\put(1798,675){\rule{1pt}{1pt}}
\put(1814,675){\rule{1pt}{1pt}}
\put(540,920){\makebox(0,0)[r]{$\gamma_z/\mu_0$}}
\put(560.0,920.0){\rule[-0.200pt]{24.090pt}{0.400pt}}
\put(221,675){\usebox{\plotpoint}}
\put(688,673.67){\rule{3.854pt}{0.400pt}}
\multiput(688.00,674.17)(8.000,-1.000){2}{\rule{1.927pt}{0.400pt}}
\put(221.0,675.0){\rule[-0.200pt]{112.500pt}{0.400pt}}
\put(849,672.67){\rule{3.854pt}{0.400pt}}
\multiput(849.00,673.17)(8.000,-1.000){2}{\rule{1.927pt}{0.400pt}}
\put(704.0,674.0){\rule[-0.200pt]{34.930pt}{0.400pt}}
\put(929,671.67){\rule{3.854pt}{0.400pt}}
\multiput(929.00,672.17)(8.000,-1.000){2}{\rule{1.927pt}{0.400pt}}
\put(865.0,673.0){\rule[-0.200pt]{15.418pt}{0.400pt}}
\put(977,670.67){\rule{3.854pt}{0.400pt}}
\multiput(977.00,671.17)(8.000,-1.000){2}{\rule{1.927pt}{0.400pt}}
\put(945.0,672.0){\rule[-0.200pt]{7.709pt}{0.400pt}}
\put(1009,669.67){\rule{4.095pt}{0.400pt}}
\multiput(1009.00,670.17)(8.500,-1.000){2}{\rule{2.048pt}{0.400pt}}
\put(993.0,671.0){\rule[-0.200pt]{3.854pt}{0.400pt}}
\put(1042,668.67){\rule{3.854pt}{0.400pt}}
\multiput(1042.00,669.17)(8.000,-1.000){2}{\rule{1.927pt}{0.400pt}}
\put(1026.0,670.0){\rule[-0.200pt]{3.854pt}{0.400pt}}
\put(1074,667.67){\rule{3.854pt}{0.400pt}}
\multiput(1074.00,668.17)(8.000,-1.000){2}{\rule{1.927pt}{0.400pt}}
\put(1090,666.67){\rule{3.854pt}{0.400pt}}
\multiput(1090.00,667.17)(8.000,-1.000){2}{\rule{1.927pt}{0.400pt}}
\put(1106,665.67){\rule{3.854pt}{0.400pt}}
\multiput(1106.00,666.17)(8.000,-1.000){2}{\rule{1.927pt}{0.400pt}}
\put(1122,664.67){\rule{3.854pt}{0.400pt}}
\multiput(1122.00,665.17)(8.000,-1.000){2}{\rule{1.927pt}{0.400pt}}
\put(1138,663.67){\rule{3.854pt}{0.400pt}}
\multiput(1138.00,664.17)(8.000,-1.000){2}{\rule{1.927pt}{0.400pt}}
\put(1154,662.67){\rule{3.854pt}{0.400pt}}
\multiput(1154.00,663.17)(8.000,-1.000){2}{\rule{1.927pt}{0.400pt}}
\put(1170,661.17){\rule{3.300pt}{0.400pt}}
\multiput(1170.00,662.17)(9.151,-2.000){2}{\rule{1.650pt}{0.400pt}}
\put(1186,659.67){\rule{4.095pt}{0.400pt}}
\multiput(1186.00,660.17)(8.500,-1.000){2}{\rule{2.048pt}{0.400pt}}
\put(1203,658.17){\rule{3.300pt}{0.400pt}}
\multiput(1203.00,659.17)(9.151,-2.000){2}{\rule{1.650pt}{0.400pt}}
\put(1219,656.17){\rule{3.300pt}{0.400pt}}
\multiput(1219.00,657.17)(9.151,-2.000){2}{\rule{1.650pt}{0.400pt}}
\put(1235,654.17){\rule{3.300pt}{0.400pt}}
\multiput(1235.00,655.17)(9.151,-2.000){2}{\rule{1.650pt}{0.400pt}}
\put(1251,652.17){\rule{3.300pt}{0.400pt}}
\multiput(1251.00,653.17)(9.151,-2.000){2}{\rule{1.650pt}{0.400pt}}
\multiput(1267.00,650.95)(3.365,-0.447){3}{\rule{2.233pt}{0.108pt}}
\multiput(1267.00,651.17)(11.365,-3.000){2}{\rule{1.117pt}{0.400pt}}
\put(1283,647.17){\rule{3.300pt}{0.400pt}}
\multiput(1283.00,648.17)(9.151,-2.000){2}{\rule{1.650pt}{0.400pt}}
\multiput(1299.00,645.95)(3.365,-0.447){3}{\rule{2.233pt}{0.108pt}}
\multiput(1299.00,646.17)(11.365,-3.000){2}{\rule{1.117pt}{0.400pt}}
\multiput(1315.00,642.94)(2.236,-0.468){5}{\rule{1.700pt}{0.113pt}}
\multiput(1315.00,643.17)(12.472,-4.000){2}{\rule{0.850pt}{0.400pt}}
\multiput(1331.00,638.95)(3.365,-0.447){3}{\rule{2.233pt}{0.108pt}}
\multiput(1331.00,639.17)(11.365,-3.000){2}{\rule{1.117pt}{0.400pt}}
\multiput(1347.00,635.94)(2.236,-0.468){5}{\rule{1.700pt}{0.113pt}}
\multiput(1347.00,636.17)(12.472,-4.000){2}{\rule{0.850pt}{0.400pt}}
\multiput(1363.00,631.93)(1.823,-0.477){7}{\rule{1.460pt}{0.115pt}}
\multiput(1363.00,632.17)(13.970,-5.000){2}{\rule{0.730pt}{0.400pt}}
\multiput(1380.00,626.94)(2.236,-0.468){5}{\rule{1.700pt}{0.113pt}}
\multiput(1380.00,627.17)(12.472,-4.000){2}{\rule{0.850pt}{0.400pt}}
\multiput(1396.00,622.93)(1.395,-0.482){9}{\rule{1.167pt}{0.116pt}}
\multiput(1396.00,623.17)(13.579,-6.000){2}{\rule{0.583pt}{0.400pt}}
\multiput(1412.00,616.93)(1.712,-0.477){7}{\rule{1.380pt}{0.115pt}}
\multiput(1412.00,617.17)(13.136,-5.000){2}{\rule{0.690pt}{0.400pt}}
\multiput(1428.00,611.93)(1.395,-0.482){9}{\rule{1.167pt}{0.116pt}}
\multiput(1428.00,612.17)(13.579,-6.000){2}{\rule{0.583pt}{0.400pt}}
\multiput(1444.00,605.93)(1.179,-0.485){11}{\rule{1.014pt}{0.117pt}}
\multiput(1444.00,606.17)(13.895,-7.000){2}{\rule{0.507pt}{0.400pt}}
\multiput(1460.00,598.93)(1.179,-0.485){11}{\rule{1.014pt}{0.117pt}}
\multiput(1460.00,599.17)(13.895,-7.000){2}{\rule{0.507pt}{0.400pt}}
\multiput(1476.00,591.93)(1.022,-0.488){13}{\rule{0.900pt}{0.117pt}}
\multiput(1476.00,592.17)(14.132,-8.000){2}{\rule{0.450pt}{0.400pt}}
\multiput(1492.00,583.93)(1.022,-0.488){13}{\rule{0.900pt}{0.117pt}}
\multiput(1492.00,584.17)(14.132,-8.000){2}{\rule{0.450pt}{0.400pt}}
\multiput(1508.00,575.93)(0.902,-0.489){15}{\rule{0.811pt}{0.118pt}}
\multiput(1508.00,576.17)(14.316,-9.000){2}{\rule{0.406pt}{0.400pt}}
\multiput(1524.00,566.92)(0.808,-0.491){17}{\rule{0.740pt}{0.118pt}}
\multiput(1524.00,567.17)(14.464,-10.000){2}{\rule{0.370pt}{0.400pt}}
\multiput(1540.00,556.92)(0.860,-0.491){17}{\rule{0.780pt}{0.118pt}}
\multiput(1540.00,557.17)(15.381,-10.000){2}{\rule{0.390pt}{0.400pt}}
\multiput(1557.00,546.92)(0.732,-0.492){19}{\rule{0.682pt}{0.118pt}}
\multiput(1557.00,547.17)(14.585,-11.000){2}{\rule{0.341pt}{0.400pt}}
\multiput(1573.00,535.92)(0.732,-0.492){19}{\rule{0.682pt}{0.118pt}}
\multiput(1573.00,536.17)(14.585,-11.000){2}{\rule{0.341pt}{0.400pt}}
\multiput(1589.00,524.92)(0.669,-0.492){21}{\rule{0.633pt}{0.119pt}}
\multiput(1589.00,525.17)(14.685,-12.000){2}{\rule{0.317pt}{0.400pt}}
\multiput(1605.00,512.92)(0.616,-0.493){23}{\rule{0.592pt}{0.119pt}}
\multiput(1605.00,513.17)(14.771,-13.000){2}{\rule{0.296pt}{0.400pt}}
\multiput(1621.00,499.92)(0.616,-0.493){23}{\rule{0.592pt}{0.119pt}}
\multiput(1621.00,500.17)(14.771,-13.000){2}{\rule{0.296pt}{0.400pt}}
\multiput(1637.00,486.92)(0.570,-0.494){25}{\rule{0.557pt}{0.119pt}}
\multiput(1637.00,487.17)(14.844,-14.000){2}{\rule{0.279pt}{0.400pt}}
\multiput(1653.00,472.92)(0.531,-0.494){27}{\rule{0.527pt}{0.119pt}}
\multiput(1653.00,473.17)(14.907,-15.000){2}{\rule{0.263pt}{0.400pt}}
\multiput(1669.00,457.92)(0.531,-0.494){27}{\rule{0.527pt}{0.119pt}}
\multiput(1669.00,458.17)(14.907,-15.000){2}{\rule{0.263pt}{0.400pt}}
\multiput(1685.00,442.92)(0.497,-0.494){29}{\rule{0.500pt}{0.119pt}}
\multiput(1685.00,443.17)(14.962,-16.000){2}{\rule{0.250pt}{0.400pt}}
\multiput(1701.58,425.82)(0.494,-0.529){29}{\rule{0.119pt}{0.525pt}}
\multiput(1700.17,426.91)(16.000,-15.910){2}{\rule{0.400pt}{0.263pt}}
\multiput(1717.00,409.92)(0.497,-0.495){31}{\rule{0.500pt}{0.119pt}}
\multiput(1717.00,410.17)(15.962,-17.000){2}{\rule{0.250pt}{0.400pt}}
\multiput(1734.58,391.82)(0.494,-0.529){29}{\rule{0.119pt}{0.525pt}}
\multiput(1733.17,392.91)(16.000,-15.910){2}{\rule{0.400pt}{0.263pt}}
\multiput(1750.58,374.72)(0.494,-0.561){29}{\rule{0.119pt}{0.550pt}}
\multiput(1749.17,375.86)(16.000,-16.858){2}{\rule{0.400pt}{0.275pt}}
\multiput(1766.58,356.61)(0.494,-0.593){29}{\rule{0.119pt}{0.575pt}}
\multiput(1765.17,357.81)(16.000,-17.807){2}{\rule{0.400pt}{0.288pt}}
\multiput(1782.58,337.61)(0.494,-0.593){29}{\rule{0.119pt}{0.575pt}}
\multiput(1781.17,338.81)(16.000,-17.807){2}{\rule{0.400pt}{0.288pt}}
\multiput(1798.58,318.61)(0.494,-0.593){29}{\rule{0.119pt}{0.575pt}}
\multiput(1797.17,319.81)(16.000,-17.807){2}{\rule{0.400pt}{0.288pt}}
\put(1058.0,669.0){\rule[-0.200pt]{3.854pt}{0.400pt}}
\end{picture}

%% file: gama-s.tex
\setlength{\unitlength}{0.240900pt}
\ifx\plotpoint\undefined\newsavebox{\plotpoint}\fi
\sbox{\plotpoint}{\rule[-0.200pt]{0.400pt}{0.400pt}}%
\begin{picture}(1875,1350)(0,0)
\font\gnuplot=cmr10 at 10pt
\gnuplot
\sbox{\plotpoint}{\rule[-0.200pt]{0.400pt}{0.400pt}}%
\put(221.0,123.0){\rule[-0.200pt]{4.818pt}{0.400pt}}
\put(201,123){\makebox(0,0)[r]{ 0.001}}
\put(1794.0,123.0){\rule[-0.200pt]{4.818pt}{0.400pt}}
\put(221.0,178.0){\rule[-0.200pt]{2.409pt}{0.400pt}}
\put(1804.0,178.0){\rule[-0.200pt]{2.409pt}{0.400pt}}
\put(221.0,252.0){\rule[-0.200pt]{2.409pt}{0.400pt}}
\put(1804.0,252.0){\rule[-0.200pt]{2.409pt}{0.400pt}}
\put(221.0,289.0){\rule[-0.200pt]{2.409pt}{0.400pt}}
\put(1804.0,289.0){\rule[-0.200pt]{2.409pt}{0.400pt}}
\put(221.0,307.0){\rule[-0.200pt]{4.818pt}{0.400pt}}
\put(201,307){\makebox(0,0)[r]{ 0.01}}
\put(1794.0,307.0){\rule[-0.200pt]{4.818pt}{0.400pt}}
\put(221.0,362.0){\rule[-0.200pt]{2.409pt}{0.400pt}}
\put(1804.0,362.0){\rule[-0.200pt]{2.409pt}{0.400pt}}
\put(221.0,436.0){\rule[-0.200pt]{2.409pt}{0.400pt}}
\put(1804.0,436.0){\rule[-0.200pt]{2.409pt}{0.400pt}}
\put(221.0,473.0){\rule[-0.200pt]{2.409pt}{0.400pt}}
\put(1804.0,473.0){\rule[-0.200pt]{2.409pt}{0.400pt}}
\put(221.0,491.0){\rule[-0.200pt]{4.818pt}{0.400pt}}
\put(201,491){\makebox(0,0)[r]{ 0.1}}
\put(1794.0,491.0){\rule[-0.200pt]{4.818pt}{0.400pt}}
\put(221.0,546.0){\rule[-0.200pt]{2.409pt}{0.400pt}}
\put(1804.0,546.0){\rule[-0.200pt]{2.409pt}{0.400pt}}
\put(221.0,620.0){\rule[-0.200pt]{2.409pt}{0.400pt}}
\put(1804.0,620.0){\rule[-0.200pt]{2.409pt}{0.400pt}}
\put(221.0,657.0){\rule[-0.200pt]{2.409pt}{0.400pt}}
\put(1804.0,657.0){\rule[-0.200pt]{2.409pt}{0.400pt}}
\put(221.0,675.0){\rule[-0.200pt]{4.818pt}{0.400pt}}
\put(201,675){\makebox(0,0)[r]{ 1}}
\put(1794.0,675.0){\rule[-0.200pt]{4.818pt}{0.400pt}}
\put(221.0,730.0){\rule[-0.200pt]{2.409pt}{0.400pt}}
\put(1804.0,730.0){\rule[-0.200pt]{2.409pt}{0.400pt}}
\put(221.0,804.0){\rule[-0.200pt]{2.409pt}{0.400pt}}
\put(1804.0,804.0){\rule[-0.200pt]{2.409pt}{0.400pt}}
\put(221.0,841.0){\rule[-0.200pt]{2.409pt}{0.400pt}}
\put(1804.0,841.0){\rule[-0.200pt]{2.409pt}{0.400pt}}
\put(221.0,859.0){\rule[-0.200pt]{4.818pt}{0.400pt}}
\put(201,859){\makebox(0,0)[r]{ 10}}
\put(1794.0,859.0){\rule[-0.200pt]{4.818pt}{0.400pt}}
\put(221.0,914.0){\rule[-0.200pt]{2.409pt}{0.400pt}}
\put(1804.0,914.0){\rule[-0.200pt]{2.409pt}{0.400pt}}
\put(221.0,988.0){\rule[-0.200pt]{2.409pt}{0.400pt}}
\put(1804.0,988.0){\rule[-0.200pt]{2.409pt}{0.400pt}}
\put(221.0,1025.0){\rule[-0.200pt]{2.409pt}{0.400pt}}
\put(1804.0,1025.0){\rule[-0.200pt]{2.409pt}{0.400pt}}
\put(221.0,1043.0){\rule[-0.200pt]{4.818pt}{0.400pt}}
\put(201,1043){\makebox(0,0)[r]{ 100}}
\put(1794.0,1043.0){\rule[-0.200pt]{4.818pt}{0.400pt}}
\put(221.0,1098.0){\rule[-0.200pt]{2.409pt}{0.400pt}}
\put(1804.0,1098.0){\rule[-0.200pt]{2.409pt}{0.400pt}}
\put(221.0,1172.0){\rule[-0.200pt]{2.409pt}{0.400pt}}
\put(1804.0,1172.0){\rule[-0.200pt]{2.409pt}{0.400pt}}
\put(221.0,1209.0){\rule[-0.200pt]{2.409pt}{0.400pt}}
\put(1804.0,1209.0){\rule[-0.200pt]{2.409pt}{0.400pt}}
\put(221.0,1227.0){\rule[-0.200pt]{4.818pt}{0.400pt}}
\put(201,1227){\makebox(0,0)[r]{ 1000}}
\put(1794.0,1227.0){\rule[-0.200pt]{4.818pt}{0.400pt}}
\put(221.0,123.0){\rule[-0.200pt]{0.400pt}{4.818pt}}
\put(221,82){\makebox(0,0){ 0.0001}}
\put(221.0,1207.0){\rule[-0.200pt]{0.400pt}{4.818pt}}
\put(323.0,123.0){\rule[-0.200pt]{0.400pt}{2.409pt}}
\put(323.0,1217.0){\rule[-0.200pt]{0.400pt}{2.409pt}}
\put(458.0,123.0){\rule[-0.200pt]{0.400pt}{2.409pt}}
\put(458.0,1217.0){\rule[-0.200pt]{0.400pt}{2.409pt}}
\put(527.0,123.0){\rule[-0.200pt]{0.400pt}{2.409pt}}
\put(527.0,1217.0){\rule[-0.200pt]{0.400pt}{2.409pt}}
\put(560.0,123.0){\rule[-0.200pt]{0.400pt}{4.818pt}}
\put(560,82){\makebox(0,0){ 0.001}}
\put(560.0,1207.0){\rule[-0.200pt]{0.400pt}{4.818pt}}
\put(662.0,123.0){\rule[-0.200pt]{0.400pt}{2.409pt}}
\put(662.0,1217.0){\rule[-0.200pt]{0.400pt}{2.409pt}}
\put(797.0,123.0){\rule[-0.200pt]{0.400pt}{2.409pt}}
\put(797.0,1217.0){\rule[-0.200pt]{0.400pt}{2.409pt}}
\put(866.0,123.0){\rule[-0.200pt]{0.400pt}{2.409pt}}
\put(866.0,1217.0){\rule[-0.200pt]{0.400pt}{2.409pt}}
\put(899.0,123.0){\rule[-0.200pt]{0.400pt}{4.818pt}}
\put(899,82){\makebox(0,0){ 0.01}}
\put(899.0,1207.0){\rule[-0.200pt]{0.400pt}{4.818pt}}
\put(1001.0,123.0){\rule[-0.200pt]{0.400pt}{2.409pt}}
\put(1001.0,1217.0){\rule[-0.200pt]{0.400pt}{2.409pt}}
\put(1136.0,123.0){\rule[-0.200pt]{0.400pt}{2.409pt}}
\put(1136.0,1217.0){\rule[-0.200pt]{0.400pt}{2.409pt}}
\put(1205.0,123.0){\rule[-0.200pt]{0.400pt}{2.409pt}}
\put(1205.0,1217.0){\rule[-0.200pt]{0.400pt}{2.409pt}}
\put(1238.0,123.0){\rule[-0.200pt]{0.400pt}{4.818pt}}
\put(1238,82){\makebox(0,0){ 0.1}}
\put(1238.0,1207.0){\rule[-0.200pt]{0.400pt}{4.818pt}}
\put(1340.0,123.0){\rule[-0.200pt]{0.400pt}{2.409pt}}
\put(1340.0,1217.0){\rule[-0.200pt]{0.400pt}{2.409pt}}
\put(1475.0,123.0){\rule[-0.200pt]{0.400pt}{2.409pt}}
\put(1475.0,1217.0){\rule[-0.200pt]{0.400pt}{2.409pt}}
\put(1544.0,123.0){\rule[-0.200pt]{0.400pt}{2.409pt}}
\put(1544.0,1217.0){\rule[-0.200pt]{0.400pt}{2.409pt}}
\put(1577.0,123.0){\rule[-0.200pt]{0.400pt}{4.818pt}}
\put(1577,82){\makebox(0,0){ 1}}
\put(1577.0,1207.0){\rule[-0.200pt]{0.400pt}{4.818pt}}
\put(1679.0,123.0){\rule[-0.200pt]{0.400pt}{2.409pt}}
\put(1679.0,1217.0){\rule[-0.200pt]{0.400pt}{2.409pt}}
\put(1814.0,123.0){\rule[-0.200pt]{0.400pt}{2.409pt}}
\put(1814.0,1217.0){\rule[-0.200pt]{0.400pt}{2.409pt}}
\put(221.0,123.0){\rule[-0.200pt]{383.754pt}{0.400pt}}
\put(1814.0,123.0){\rule[-0.200pt]{0.400pt}{265.954pt}}
\put(221.0,1227.0){\rule[-0.200pt]{383.754pt}{0.400pt}}
\put(40,675){\makebox(0,0){$\gamma^\ast / \mu_0 $}}
\put(1017,21){\makebox(0,0){$z$}}
\put(1017,1289){\makebox(0,0){ }}
\put(221.0,123.0){\rule[-0.200pt]{0.400pt}{265.954pt}}
\put(540,1043){\makebox(0,0)[r]{$\gamma^\ast_{\ssty L}/\mu_0$}}
\multiput(560,1043)(20.756,0.000){5}{\usebox{\plotpoint}}
\put(660,1043){\usebox{\plotpoint}}
\put(221,675){\usebox{\plotpoint}}
\put(221.00,675.00){\usebox{\plotpoint}}
\put(241.76,675.00){\usebox{\plotpoint}}
\put(262.51,675.00){\usebox{\plotpoint}}
\put(283.27,675.00){\usebox{\plotpoint}}
\put(304.02,675.00){\usebox{\plotpoint}}
\put(324.78,675.00){\usebox{\plotpoint}}
\put(345.53,675.00){\usebox{\plotpoint}}
\put(366.29,675.00){\usebox{\plotpoint}}
\put(387.04,675.00){\usebox{\plotpoint}}
\put(407.80,675.00){\usebox{\plotpoint}}
\put(428.55,675.00){\usebox{\plotpoint}}
\put(449.31,675.00){\usebox{\plotpoint}}
\put(470.07,675.00){\usebox{\plotpoint}}
\put(490.82,675.00){\usebox{\plotpoint}}
\put(511.58,675.00){\usebox{\plotpoint}}
\put(532.33,675.00){\usebox{\plotpoint}}
\put(553.09,675.00){\usebox{\plotpoint}}
\put(573.84,675.00){\usebox{\plotpoint}}
\put(594.60,675.00){\usebox{\plotpoint}}
\put(615.35,675.00){\usebox{\plotpoint}}
\put(636.11,675.00){\usebox{\plotpoint}}
\put(656.86,674.89){\usebox{\plotpoint}}
\put(677.59,674.00){\usebox{\plotpoint}}
\put(698.35,674.00){\usebox{\plotpoint}}
\put(719.10,674.00){\usebox{\plotpoint}}
\put(739.86,674.00){\usebox{\plotpoint}}
\put(760.61,674.00){\usebox{\plotpoint}}
\put(781.37,674.00){\usebox{\plotpoint}}
\put(802.12,674.00){\usebox{\plotpoint}}
\put(822.87,673.57){\usebox{\plotpoint}}
\put(843.60,673.00){\usebox{\plotpoint}}
\put(864.36,673.00){\usebox{\plotpoint}}
\put(885.12,673.00){\usebox{\plotpoint}}
\put(905.85,672.45){\usebox{\plotpoint}}
\put(926.59,672.00){\usebox{\plotpoint}}
\put(947.35,671.85){\usebox{\plotpoint}}
\put(968.07,671.00){\usebox{\plotpoint}}
\put(988.81,670.26){\usebox{\plotpoint}}
\put(1009.55,669.97){\usebox{\plotpoint}}
\put(1030.28,669.00){\usebox{\plotpoint}}
\put(1051.02,668.44){\usebox{\plotpoint}}
\put(1071.73,667.14){\usebox{\plotpoint}}
\put(1092.45,665.85){\usebox{\plotpoint}}
\put(1113.16,664.55){\usebox{\plotpoint}}
\put(1133.88,663.26){\usebox{\plotpoint}}
\put(1154.59,661.93){\usebox{\plotpoint}}
\put(1175.22,659.67){\usebox{\plotpoint}}
\put(1195.88,657.84){\usebox{\plotpoint}}
\put(1216.48,655.31){\usebox{\plotpoint}}
\put(1237.06,652.61){\usebox{\plotpoint}}
\put(1257.46,648.79){\usebox{\plotpoint}}
\put(1277.86,644.96){\usebox{\plotpoint}}
\put(1298.26,641.14){\usebox{\plotpoint}}
\put(1318.61,637.10){\usebox{\plotpoint}}
\put(1338.62,631.62){\usebox{\plotpoint}}
\put(1358.62,626.09){\usebox{\plotpoint}}
\put(1378.32,619.59){\usebox{\plotpoint}}
\put(1398.02,613.12){\usebox{\plotpoint}}
\put(1417.15,605.07){\usebox{\plotpoint}}
\put(1436.40,597.33){\usebox{\plotpoint}}
\put(1455.14,588.43){\usebox{\plotpoint}}
\put(1473.36,578.49){\usebox{\plotpoint}}
\put(1491.45,568.31){\usebox{\plotpoint}}
\put(1509.06,557.34){\usebox{\plotpoint}}
\put(1526.59,546.22){\usebox{\plotpoint}}
\put(1543.66,534.42){\usebox{\plotpoint}}
\put(1560.43,522.21){\usebox{\plotpoint}}
\put(1576.54,509.12){\usebox{\plotpoint}}
\put(1592.54,495.90){\usebox{\plotpoint}}
\put(1608.06,482.13){\usebox{\plotpoint}}
\put(1623.21,467.93){\usebox{\plotpoint}}
\multiput(1637,455)(14.676,-14.676){2}{\usebox{\plotpoint}}
\put(1667.21,423.90){\usebox{\plotpoint}}
\put(1681.43,408.79){\usebox{\plotpoint}}
\put(1695.33,393.38){\usebox{\plotpoint}}
\put(1708.87,377.65){\usebox{\plotpoint}}
\put(1722.43,361.93){\usebox{\plotpoint}}
\multiput(1734,349)(12.966,-16.207){2}{\usebox{\plotpoint}}
\put(1762.06,313.93){\usebox{\plotpoint}}
\put(1774.75,297.51){\usebox{\plotpoint}}
\put(1787.33,281.00){\usebox{\plotpoint}}
\multiput(1798,267)(12.208,-16.786){2}{\usebox{\plotpoint}}
\put(1814,245){\usebox{\plotpoint}}
\put(221,675){\circle{18}}
\put(237,675){\circle{18}}
\put(253,675){\circle{18}}
\put(269,675){\circle{18}}
\put(285,675){\circle{18}}
\put(301,675){\circle{18}}
\put(318,675){\circle{18}}
\put(334,675){\circle{18}}
\put(350,675){\circle{18}}
\put(366,675){\circle{18}}
\put(382,675){\circle{18}}
\put(398,675){\circle{18}}
\put(414,675){\circle{18}}
\put(430,675){\circle{18}}
\put(446,675){\circle{18}}
\put(462,675){\circle{18}}
\put(478,675){\circle{18}}
\put(495,675){\circle{18}}
\put(511,675){\circle{18}}
\put(527,675){\circle{18}}
\put(543,675){\circle{18}}
\put(559,675){\circle{18}}
\put(575,675){\circle{18}}
\put(591,675){\circle{18}}
\put(607,675){\circle{18}}
\put(623,675){\circle{18}}
\put(639,675){\circle{18}}
\put(655,675){\circle{18}}
\put(672,674){\circle{18}}
\put(688,674){\circle{18}}
\put(704,674){\circle{18}}
\put(720,674){\circle{18}}
\put(736,674){\circle{18}}
\put(752,674){\circle{18}}
\put(768,674){\circle{18}}
\put(784,674){\circle{18}}
\put(800,674){\circle{18}}
\put(816,674){\circle{18}}
\put(832,673){\circle{18}}
\put(849,673){\circle{18}}
\put(865,673){\circle{18}}
\put(881,673){\circle{18}}
\put(897,673){\circle{18}}
\put(913,672){\circle{18}}
\put(929,672){\circle{18}}
\put(945,672){\circle{18}}
\put(961,671){\circle{18}}
\put(977,671){\circle{18}}
\put(993,670){\circle{18}}
\put(1009,670){\circle{18}}
\put(1026,669){\circle{18}}
\put(1042,669){\circle{18}}
\put(1058,668){\circle{18}}
\put(1074,667){\circle{18}}
\put(1090,666){\circle{18}}
\put(1106,665){\circle{18}}
\put(1122,664){\circle{18}}
\put(1138,663){\circle{18}}
\put(1154,662){\circle{18}}
\put(1170,660){\circle{18}}
\put(1186,659){\circle{18}}
\put(1203,657){\circle{18}}
\put(1219,655){\circle{18}}
\put(1235,653){\circle{18}}
\put(1251,650){\circle{18}}
\put(1267,647){\circle{18}}
\put(1283,644){\circle{18}}
\put(1299,641){\circle{18}}
\put(1315,638){\circle{18}}
\put(1331,634){\circle{18}}
\put(1347,629){\circle{18}}
\put(1363,625){\circle{18}}
\put(1380,619){\circle{18}}
\put(1396,614){\circle{18}}
\put(1412,607){\circle{18}}
\put(1428,601){\circle{18}}
\put(1444,594){\circle{18}}
\put(1460,586){\circle{18}}
\put(1476,577){\circle{18}}
\put(1492,568){\circle{18}}
\put(1508,558){\circle{18}}
\put(1524,548){\circle{18}}
\put(1540,537){\circle{18}}
\put(1557,525){\circle{18}}
\put(1573,512){\circle{18}}
\put(1589,499){\circle{18}}
\put(1605,485){\circle{18}}
\put(1621,470){\circle{18}}
\put(1637,455){\circle{18}}
\put(1653,439){\circle{18}}
\put(1669,422){\circle{18}}
\put(1685,405){\circle{18}}
\put(1701,387){\circle{18}}
\put(1717,368){\circle{18}}
\put(1734,349){\circle{18}}
\put(1750,329){\circle{18}}
\put(1766,309){\circle{18}}
\put(1782,288){\circle{18}}
\put(1798,267){\circle{18}}
\put(1814,245){\circle{18}}
\put(610,1043){\circle{18}}
\sbox{\plotpoint}{\rule[-0.400pt]{0.800pt}{0.800pt}}%
\put(540,1002){\makebox(0,0)[r]{$\gamma^\ast_{\ssty A}/\mu_0$}}
\put(560.0,1002.0){\rule[-0.400pt]{24.090pt}{0.800pt}}
\put(221,675){\usebox{\plotpoint}}
\put(655,673.84){\rule{4.095pt}{0.800pt}}
\multiput(655.00,673.34)(8.500,1.000){2}{\rule{2.048pt}{0.800pt}}
\put(221.0,675.0){\rule[-0.400pt]{104.551pt}{0.800pt}}
\put(816,674.84){\rule{3.854pt}{0.800pt}}
\multiput(816.00,674.34)(8.000,1.000){2}{\rule{1.927pt}{0.800pt}}
\put(672.0,676.0){\rule[-0.400pt]{34.690pt}{0.800pt}}
\put(897,675.84){\rule{3.854pt}{0.800pt}}
\multiput(897.00,675.34)(8.000,1.000){2}{\rule{1.927pt}{0.800pt}}
\put(832.0,677.0){\rule[-0.400pt]{15.658pt}{0.800pt}}
\put(945,676.84){\rule{3.854pt}{0.800pt}}
\multiput(945.00,676.34)(8.000,1.000){2}{\rule{1.927pt}{0.800pt}}
\put(913.0,678.0){\rule[-0.400pt]{7.709pt}{0.800pt}}
\put(977,677.84){\rule{3.854pt}{0.800pt}}
\multiput(977.00,677.34)(8.000,1.000){2}{\rule{1.927pt}{0.800pt}}
\put(961.0,679.0){\rule[-0.400pt]{3.854pt}{0.800pt}}
\put(1009,678.84){\rule{4.095pt}{0.800pt}}
\multiput(1009.00,678.34)(8.500,1.000){2}{\rule{2.048pt}{0.800pt}}
\put(993.0,680.0){\rule[-0.400pt]{3.854pt}{0.800pt}}
\put(1042,679.84){\rule{3.854pt}{0.800pt}}
\multiput(1042.00,679.34)(8.000,1.000){2}{\rule{1.927pt}{0.800pt}}
\put(1058,680.84){\rule{3.854pt}{0.800pt}}
\multiput(1058.00,680.34)(8.000,1.000){2}{\rule{1.927pt}{0.800pt}}
\put(1074,681.84){\rule{3.854pt}{0.800pt}}
\multiput(1074.00,681.34)(8.000,1.000){2}{\rule{1.927pt}{0.800pt}}
\put(1090,682.84){\rule{3.854pt}{0.800pt}}
\multiput(1090.00,682.34)(8.000,1.000){2}{\rule{1.927pt}{0.800pt}}
\put(1106,683.84){\rule{3.854pt}{0.800pt}}
\multiput(1106.00,683.34)(8.000,1.000){2}{\rule{1.927pt}{0.800pt}}
\put(1122,684.84){\rule{3.854pt}{0.800pt}}
\multiput(1122.00,684.34)(8.000,1.000){2}{\rule{1.927pt}{0.800pt}}
\put(1138,685.84){\rule{3.854pt}{0.800pt}}
\multiput(1138.00,685.34)(8.000,1.000){2}{\rule{1.927pt}{0.800pt}}
\put(1154,687.34){\rule{3.854pt}{0.800pt}}
\multiput(1154.00,686.34)(8.000,2.000){2}{\rule{1.927pt}{0.800pt}}
\put(1170,688.84){\rule{3.854pt}{0.800pt}}
\multiput(1170.00,688.34)(8.000,1.000){2}{\rule{1.927pt}{0.800pt}}
\put(1186,690.34){\rule{4.095pt}{0.800pt}}
\multiput(1186.00,689.34)(8.500,2.000){2}{\rule{2.048pt}{0.800pt}}
\put(1203,692.34){\rule{3.854pt}{0.800pt}}
\multiput(1203.00,691.34)(8.000,2.000){2}{\rule{1.927pt}{0.800pt}}
\put(1219,694.34){\rule{3.854pt}{0.800pt}}
\multiput(1219.00,693.34)(8.000,2.000){2}{\rule{1.927pt}{0.800pt}}
\put(1235,696.84){\rule{3.854pt}{0.800pt}}
\multiput(1235.00,695.34)(8.000,3.000){2}{\rule{1.927pt}{0.800pt}}
\put(1251,699.84){\rule{3.854pt}{0.800pt}}
\multiput(1251.00,698.34)(8.000,3.000){2}{\rule{1.927pt}{0.800pt}}
\put(1267,702.84){\rule{3.854pt}{0.800pt}}
\multiput(1267.00,701.34)(8.000,3.000){2}{\rule{1.927pt}{0.800pt}}
\put(1283,705.84){\rule{3.854pt}{0.800pt}}
\multiput(1283.00,704.34)(8.000,3.000){2}{\rule{1.927pt}{0.800pt}}
\put(1299,708.84){\rule{3.854pt}{0.800pt}}
\multiput(1299.00,707.34)(8.000,3.000){2}{\rule{1.927pt}{0.800pt}}
\put(1315,712.34){\rule{3.400pt}{0.800pt}}
\multiput(1315.00,710.34)(8.943,4.000){2}{\rule{1.700pt}{0.800pt}}
\multiput(1331.00,717.38)(2.271,0.560){3}{\rule{2.760pt}{0.135pt}}
\multiput(1331.00,714.34)(10.271,5.000){2}{\rule{1.380pt}{0.800pt}}
\put(1347,721.34){\rule{3.400pt}{0.800pt}}
\multiput(1347.00,719.34)(8.943,4.000){2}{\rule{1.700pt}{0.800pt}}
\multiput(1363.00,726.39)(1.690,0.536){5}{\rule{2.467pt}{0.129pt}}
\multiput(1363.00,723.34)(11.880,6.000){2}{\rule{1.233pt}{0.800pt}}
\multiput(1380.00,732.38)(2.271,0.560){3}{\rule{2.760pt}{0.135pt}}
\multiput(1380.00,729.34)(10.271,5.000){2}{\rule{1.380pt}{0.800pt}}
\multiput(1396.00,737.40)(1.263,0.526){7}{\rule{2.029pt}{0.127pt}}
\multiput(1396.00,734.34)(11.790,7.000){2}{\rule{1.014pt}{0.800pt}}
\multiput(1412.00,744.39)(1.579,0.536){5}{\rule{2.333pt}{0.129pt}}
\multiput(1412.00,741.34)(11.157,6.000){2}{\rule{1.167pt}{0.800pt}}
\multiput(1428.00,750.40)(1.263,0.526){7}{\rule{2.029pt}{0.127pt}}
\multiput(1428.00,747.34)(11.790,7.000){2}{\rule{1.014pt}{0.800pt}}
\multiput(1444.00,757.40)(1.066,0.520){9}{\rule{1.800pt}{0.125pt}}
\multiput(1444.00,754.34)(12.264,8.000){2}{\rule{0.900pt}{0.800pt}}
\multiput(1460.00,765.40)(0.927,0.516){11}{\rule{1.622pt}{0.124pt}}
\multiput(1460.00,762.34)(12.633,9.000){2}{\rule{0.811pt}{0.800pt}}
\multiput(1476.00,774.40)(0.927,0.516){11}{\rule{1.622pt}{0.124pt}}
\multiput(1476.00,771.34)(12.633,9.000){2}{\rule{0.811pt}{0.800pt}}
\multiput(1492.00,783.40)(0.821,0.514){13}{\rule{1.480pt}{0.124pt}}
\multiput(1492.00,780.34)(12.928,10.000){2}{\rule{0.740pt}{0.800pt}}
\multiput(1508.00,793.40)(0.821,0.514){13}{\rule{1.480pt}{0.124pt}}
\multiput(1508.00,790.34)(12.928,10.000){2}{\rule{0.740pt}{0.800pt}}
\multiput(1524.00,803.40)(0.739,0.512){15}{\rule{1.364pt}{0.123pt}}
\multiput(1524.00,800.34)(13.170,11.000){2}{\rule{0.682pt}{0.800pt}}
\multiput(1540.00,814.41)(0.717,0.511){17}{\rule{1.333pt}{0.123pt}}
\multiput(1540.00,811.34)(14.233,12.000){2}{\rule{0.667pt}{0.800pt}}
\multiput(1557.00,826.41)(0.616,0.509){19}{\rule{1.185pt}{0.123pt}}
\multiput(1557.00,823.34)(13.541,13.000){2}{\rule{0.592pt}{0.800pt}}
\multiput(1573.00,839.41)(0.616,0.509){19}{\rule{1.185pt}{0.123pt}}
\multiput(1573.00,836.34)(13.541,13.000){2}{\rule{0.592pt}{0.800pt}}
\multiput(1589.00,852.41)(0.569,0.509){21}{\rule{1.114pt}{0.123pt}}
\multiput(1589.00,849.34)(13.687,14.000){2}{\rule{0.557pt}{0.800pt}}
\multiput(1605.00,866.41)(0.529,0.508){23}{\rule{1.053pt}{0.122pt}}
\multiput(1605.00,863.34)(13.814,15.000){2}{\rule{0.527pt}{0.800pt}}
\multiput(1621.00,881.41)(0.529,0.508){23}{\rule{1.053pt}{0.122pt}}
\multiput(1621.00,878.34)(13.814,15.000){2}{\rule{0.527pt}{0.800pt}}
\multiput(1637.00,896.41)(0.494,0.507){25}{\rule{1.000pt}{0.122pt}}
\multiput(1637.00,893.34)(13.924,16.000){2}{\rule{0.500pt}{0.800pt}}
\multiput(1654.41,911.00)(0.507,0.527){25}{\rule{0.122pt}{1.050pt}}
\multiput(1651.34,911.00)(16.000,14.821){2}{\rule{0.800pt}{0.525pt}}
\multiput(1670.41,928.00)(0.507,0.527){25}{\rule{0.122pt}{1.050pt}}
\multiput(1667.34,928.00)(16.000,14.821){2}{\rule{0.800pt}{0.525pt}}
\multiput(1686.41,945.00)(0.507,0.560){25}{\rule{0.122pt}{1.100pt}}
\multiput(1683.34,945.00)(16.000,15.717){2}{\rule{0.800pt}{0.550pt}}
\multiput(1702.41,963.00)(0.507,0.593){25}{\rule{0.122pt}{1.150pt}}
\multiput(1699.34,963.00)(16.000,16.613){2}{\rule{0.800pt}{0.575pt}}
\multiput(1718.41,982.00)(0.507,0.556){27}{\rule{0.122pt}{1.094pt}}
\multiput(1715.34,982.00)(17.000,16.729){2}{\rule{0.800pt}{0.547pt}}
\multiput(1735.41,1001.00)(0.507,0.626){25}{\rule{0.122pt}{1.200pt}}
\multiput(1732.34,1001.00)(16.000,17.509){2}{\rule{0.800pt}{0.600pt}}
\multiput(1751.41,1021.00)(0.507,0.626){25}{\rule{0.122pt}{1.200pt}}
\multiput(1748.34,1021.00)(16.000,17.509){2}{\rule{0.800pt}{0.600pt}}
\multiput(1767.41,1041.00)(0.507,0.659){25}{\rule{0.122pt}{1.250pt}}
\multiput(1764.34,1041.00)(16.000,18.406){2}{\rule{0.800pt}{0.625pt}}
\multiput(1783.41,1062.00)(0.507,0.659){25}{\rule{0.122pt}{1.250pt}}
\multiput(1780.34,1062.00)(16.000,18.406){2}{\rule{0.800pt}{0.625pt}}
\multiput(1799.41,1083.00)(0.507,0.692){25}{\rule{0.122pt}{1.300pt}}
\multiput(1796.34,1083.00)(16.000,19.302){2}{\rule{0.800pt}{0.650pt}}
\put(1026.0,681.0){\rule[-0.400pt]{3.854pt}{0.800pt}}
\sbox{\plotpoint}{\rule[-0.200pt]{0.400pt}{0.400pt}}%
\put(540,961){\makebox(0,0)[r]{$\gamma^\ast_{\ssty G}/\mu_0$}}
\put(610,961){\rule{1pt}{1pt}}
\put(221,675){\rule{1pt}{1pt}}
\put(237,675){\rule{1pt}{1pt}}
\put(253,675){\rule{1pt}{1pt}}
\put(269,675){\rule{1pt}{1pt}}
\put(285,675){\rule{1pt}{1pt}}
\put(301,675){\rule{1pt}{1pt}}
\put(318,675){\rule{1pt}{1pt}}
\put(334,675){\rule{1pt}{1pt}}
\put(350,675){\rule{1pt}{1pt}}
\put(366,675){\rule{1pt}{1pt}}
\put(382,675){\rule{1pt}{1pt}}
\put(398,675){\rule{1pt}{1pt}}
\put(414,675){\rule{1pt}{1pt}}
\put(430,675){\rule{1pt}{1pt}}
\put(446,675){\rule{1pt}{1pt}}
\put(462,675){\rule{1pt}{1pt}}
\put(478,675){\rule{1pt}{1pt}}
\put(495,675){\rule{1pt}{1pt}}
\put(511,675){\rule{1pt}{1pt}}
\put(527,675){\rule{1pt}{1pt}}
\put(543,675){\rule{1pt}{1pt}}
\put(559,675){\rule{1pt}{1pt}}
\put(575,675){\rule{1pt}{1pt}}
\put(591,675){\rule{1pt}{1pt}}
\put(607,675){\rule{1pt}{1pt}}
\put(623,675){\rule{1pt}{1pt}}
\put(639,675){\rule{1pt}{1pt}}
\put(655,675){\rule{1pt}{1pt}}
\put(672,675){\rule{1pt}{1pt}}
\put(688,675){\rule{1pt}{1pt}}
\put(704,675){\rule{1pt}{1pt}}
\put(720,675){\rule{1pt}{1pt}}
\put(736,675){\rule{1pt}{1pt}}
\put(752,675){\rule{1pt}{1pt}}
\put(768,675){\rule{1pt}{1pt}}
\put(784,675){\rule{1pt}{1pt}}
\put(800,675){\rule{1pt}{1pt}}
\put(816,675){\rule{1pt}{1pt}}
\put(832,675){\rule{1pt}{1pt}}
\put(849,675){\rule{1pt}{1pt}}
\put(865,675){\rule{1pt}{1pt}}
\put(881,675){\rule{1pt}{1pt}}
\put(897,675){\rule{1pt}{1pt}}
\put(913,675){\rule{1pt}{1pt}}
\put(929,675){\rule{1pt}{1pt}}
\put(945,675){\rule{1pt}{1pt}}
\put(961,675){\rule{1pt}{1pt}}
\put(977,675){\rule{1pt}{1pt}}
\put(993,675){\rule{1pt}{1pt}}
\put(1009,675){\rule{1pt}{1pt}}
\put(1026,675){\rule{1pt}{1pt}}
\put(1042,675){\rule{1pt}{1pt}}
\put(1058,675){\rule{1pt}{1pt}}
\put(1074,675){\rule{1pt}{1pt}}
\put(1090,675){\rule{1pt}{1pt}}
\put(1106,675){\rule{1pt}{1pt}}
\put(1122,675){\rule{1pt}{1pt}}
\put(1138,675){\rule{1pt}{1pt}}
\put(1154,675){\rule{1pt}{1pt}}
\put(1170,675){\rule{1pt}{1pt}}
\put(1186,675){\rule{1pt}{1pt}}
\put(1203,675){\rule{1pt}{1pt}}
\put(1219,675){\rule{1pt}{1pt}}
\put(1235,675){\rule{1pt}{1pt}}
\put(1251,675){\rule{1pt}{1pt}}
\put(1267,675){\rule{1pt}{1pt}}
\put(1283,675){\rule{1pt}{1pt}}
\put(1299,675){\rule{1pt}{1pt}}
\put(1315,675){\rule{1pt}{1pt}}
\put(1331,675){\rule{1pt}{1pt}}
\put(1347,675){\rule{1pt}{1pt}}
\put(1363,675){\rule{1pt}{1pt}}
\put(1380,675){\rule{1pt}{1pt}}
\put(1396,675){\rule{1pt}{1pt}}
\put(1412,675){\rule{1pt}{1pt}}
\put(1428,675){\rule{1pt}{1pt}}
\put(1444,675){\rule{1pt}{1pt}}
\put(1460,675){\rule{1pt}{1pt}}
\put(1476,675){\rule{1pt}{1pt}}
\put(1492,675){\rule{1pt}{1pt}}
\put(1508,675){\rule{1pt}{1pt}}
\put(1524,675){\rule{1pt}{1pt}}
\put(1540,675){\rule{1pt}{1pt}}
\put(1557,675){\rule{1pt}{1pt}}
\put(1573,675){\rule{1pt}{1pt}}
\put(1589,675){\rule{1pt}{1pt}}
\put(1605,675){\rule{1pt}{1pt}}
\put(1621,675){\rule{1pt}{1pt}}
\put(1637,675){\rule{1pt}{1pt}}
\put(1653,675){\rule{1pt}{1pt}}
\put(1669,675){\rule{1pt}{1pt}}
\put(1685,675){\rule{1pt}{1pt}}
\put(1701,675){\rule{1pt}{1pt}}
\put(1717,675){\rule{1pt}{1pt}}
\put(1734,675){\rule{1pt}{1pt}}
\put(1750,675){\rule{1pt}{1pt}}
\put(1766,675){\rule{1pt}{1pt}}
\put(1782,675){\rule{1pt}{1pt}}
\put(1798,675){\rule{1pt}{1pt}}
\put(1814,675){\rule{1pt}{1pt}}
\put(540,920){\makebox(0,0)[r]{$\gamma^\ast_z/\mu_0$}}
\put(560.0,920.0){\rule[-0.200pt]{24.090pt}{0.400pt}}
\put(221,675){\usebox{\plotpoint}}
\put(704,673.67){\rule{3.854pt}{0.400pt}}
\multiput(704.00,674.17)(8.000,-1.000){2}{\rule{1.927pt}{0.400pt}}
\put(221.0,675.0){\rule[-0.200pt]{116.355pt}{0.400pt}}
\put(865,672.67){\rule{3.854pt}{0.400pt}}
\multiput(865.00,673.17)(8.000,-1.000){2}{\rule{1.927pt}{0.400pt}}
\put(720.0,674.0){\rule[-0.200pt]{34.930pt}{0.400pt}}
\put(945,671.67){\rule{3.854pt}{0.400pt}}
\multiput(945.00,672.17)(8.000,-1.000){2}{\rule{1.927pt}{0.400pt}}
\put(881.0,673.0){\rule[-0.200pt]{15.418pt}{0.400pt}}
\put(993,670.67){\rule{3.854pt}{0.400pt}}
\multiput(993.00,671.17)(8.000,-1.000){2}{\rule{1.927pt}{0.400pt}}
\put(961.0,672.0){\rule[-0.200pt]{7.709pt}{0.400pt}}
\put(1026,669.67){\rule{3.854pt}{0.400pt}}
\multiput(1026.00,670.17)(8.000,-1.000){2}{\rule{1.927pt}{0.400pt}}
\put(1009.0,671.0){\rule[-0.200pt]{4.095pt}{0.400pt}}
\put(1058,668.67){\rule{3.854pt}{0.400pt}}
\multiput(1058.00,669.17)(8.000,-1.000){2}{\rule{1.927pt}{0.400pt}}
\put(1042.0,670.0){\rule[-0.200pt]{3.854pt}{0.400pt}}
\put(1090,667.67){\rule{3.854pt}{0.400pt}}
\multiput(1090.00,668.17)(8.000,-1.000){2}{\rule{1.927pt}{0.400pt}}
\put(1106,666.67){\rule{3.854pt}{0.400pt}}
\multiput(1106.00,667.17)(8.000,-1.000){2}{\rule{1.927pt}{0.400pt}}
\put(1122,665.67){\rule{3.854pt}{0.400pt}}
\multiput(1122.00,666.17)(8.000,-1.000){2}{\rule{1.927pt}{0.400pt}}
\put(1138,664.67){\rule{3.854pt}{0.400pt}}
\multiput(1138.00,665.17)(8.000,-1.000){2}{\rule{1.927pt}{0.400pt}}
\put(1154,663.67){\rule{3.854pt}{0.400pt}}
\multiput(1154.00,664.17)(8.000,-1.000){2}{\rule{1.927pt}{0.400pt}}
\put(1170,662.67){\rule{3.854pt}{0.400pt}}
\multiput(1170.00,663.17)(8.000,-1.000){2}{\rule{1.927pt}{0.400pt}}
\put(1186,661.17){\rule{3.500pt}{0.400pt}}
\multiput(1186.00,662.17)(9.736,-2.000){2}{\rule{1.750pt}{0.400pt}}
\put(1203,659.67){\rule{3.854pt}{0.400pt}}
\multiput(1203.00,660.17)(8.000,-1.000){2}{\rule{1.927pt}{0.400pt}}
\put(1219,658.17){\rule{3.300pt}{0.400pt}}
\multiput(1219.00,659.17)(9.151,-2.000){2}{\rule{1.650pt}{0.400pt}}
\put(1235,656.17){\rule{3.300pt}{0.400pt}}
\multiput(1235.00,657.17)(9.151,-2.000){2}{\rule{1.650pt}{0.400pt}}
\put(1251,654.17){\rule{3.300pt}{0.400pt}}
\multiput(1251.00,655.17)(9.151,-2.000){2}{\rule{1.650pt}{0.400pt}}
\put(1267,652.17){\rule{3.300pt}{0.400pt}}
\multiput(1267.00,653.17)(9.151,-2.000){2}{\rule{1.650pt}{0.400pt}}
\put(1283,650.17){\rule{3.300pt}{0.400pt}}
\multiput(1283.00,651.17)(9.151,-2.000){2}{\rule{1.650pt}{0.400pt}}
\multiput(1299.00,648.95)(3.365,-0.447){3}{\rule{2.233pt}{0.108pt}}
\multiput(1299.00,649.17)(11.365,-3.000){2}{\rule{1.117pt}{0.400pt}}
\multiput(1315.00,645.95)(3.365,-0.447){3}{\rule{2.233pt}{0.108pt}}
\multiput(1315.00,646.17)(11.365,-3.000){2}{\rule{1.117pt}{0.400pt}}
\multiput(1331.00,642.94)(2.236,-0.468){5}{\rule{1.700pt}{0.113pt}}
\multiput(1331.00,643.17)(12.472,-4.000){2}{\rule{0.850pt}{0.400pt}}
\multiput(1347.00,638.95)(3.365,-0.447){3}{\rule{2.233pt}{0.108pt}}
\multiput(1347.00,639.17)(11.365,-3.000){2}{\rule{1.117pt}{0.400pt}}
\multiput(1363.00,635.94)(2.382,-0.468){5}{\rule{1.800pt}{0.113pt}}
\multiput(1363.00,636.17)(13.264,-4.000){2}{\rule{0.900pt}{0.400pt}}
\multiput(1380.00,631.93)(1.712,-0.477){7}{\rule{1.380pt}{0.115pt}}
\multiput(1380.00,632.17)(13.136,-5.000){2}{\rule{0.690pt}{0.400pt}}
\multiput(1396.00,626.94)(2.236,-0.468){5}{\rule{1.700pt}{0.113pt}}
\multiput(1396.00,627.17)(12.472,-4.000){2}{\rule{0.850pt}{0.400pt}}
\multiput(1412.00,622.93)(1.712,-0.477){7}{\rule{1.380pt}{0.115pt}}
\multiput(1412.00,623.17)(13.136,-5.000){2}{\rule{0.690pt}{0.400pt}}
\multiput(1428.00,617.93)(1.395,-0.482){9}{\rule{1.167pt}{0.116pt}}
\multiput(1428.00,618.17)(13.579,-6.000){2}{\rule{0.583pt}{0.400pt}}
\multiput(1444.00,611.93)(1.395,-0.482){9}{\rule{1.167pt}{0.116pt}}
\multiput(1444.00,612.17)(13.579,-6.000){2}{\rule{0.583pt}{0.400pt}}
\multiput(1460.00,605.93)(1.179,-0.485){11}{\rule{1.014pt}{0.117pt}}
\multiput(1460.00,606.17)(13.895,-7.000){2}{\rule{0.507pt}{0.400pt}}
\multiput(1476.00,598.93)(1.179,-0.485){11}{\rule{1.014pt}{0.117pt}}
\multiput(1476.00,599.17)(13.895,-7.000){2}{\rule{0.507pt}{0.400pt}}
\multiput(1492.00,591.93)(1.179,-0.485){11}{\rule{1.014pt}{0.117pt}}
\multiput(1492.00,592.17)(13.895,-7.000){2}{\rule{0.507pt}{0.400pt}}
\multiput(1508.00,584.93)(1.022,-0.488){13}{\rule{0.900pt}{0.117pt}}
\multiput(1508.00,585.17)(14.132,-8.000){2}{\rule{0.450pt}{0.400pt}}
\multiput(1524.00,576.93)(0.902,-0.489){15}{\rule{0.811pt}{0.118pt}}
\multiput(1524.00,577.17)(14.316,-9.000){2}{\rule{0.406pt}{0.400pt}}
\multiput(1540.00,567.93)(0.961,-0.489){15}{\rule{0.856pt}{0.118pt}}
\multiput(1540.00,568.17)(15.224,-9.000){2}{\rule{0.428pt}{0.400pt}}
\multiput(1557.00,558.92)(0.808,-0.491){17}{\rule{0.740pt}{0.118pt}}
\multiput(1557.00,559.17)(14.464,-10.000){2}{\rule{0.370pt}{0.400pt}}
\multiput(1573.00,548.92)(0.732,-0.492){19}{\rule{0.682pt}{0.118pt}}
\multiput(1573.00,549.17)(14.585,-11.000){2}{\rule{0.341pt}{0.400pt}}
\multiput(1589.00,537.92)(0.732,-0.492){19}{\rule{0.682pt}{0.118pt}}
\multiput(1589.00,538.17)(14.585,-11.000){2}{\rule{0.341pt}{0.400pt}}
\multiput(1605.00,526.92)(0.669,-0.492){21}{\rule{0.633pt}{0.119pt}}
\multiput(1605.00,527.17)(14.685,-12.000){2}{\rule{0.317pt}{0.400pt}}
\multiput(1621.00,514.92)(0.669,-0.492){21}{\rule{0.633pt}{0.119pt}}
\multiput(1621.00,515.17)(14.685,-12.000){2}{\rule{0.317pt}{0.400pt}}
\multiput(1637.00,502.92)(0.616,-0.493){23}{\rule{0.592pt}{0.119pt}}
\multiput(1637.00,503.17)(14.771,-13.000){2}{\rule{0.296pt}{0.400pt}}
\multiput(1653.00,489.92)(0.570,-0.494){25}{\rule{0.557pt}{0.119pt}}
\multiput(1653.00,490.17)(14.844,-14.000){2}{\rule{0.279pt}{0.400pt}}
\multiput(1669.00,475.92)(0.570,-0.494){25}{\rule{0.557pt}{0.119pt}}
\multiput(1669.00,476.17)(14.844,-14.000){2}{\rule{0.279pt}{0.400pt}}
\multiput(1685.00,461.92)(0.531,-0.494){27}{\rule{0.527pt}{0.119pt}}
\multiput(1685.00,462.17)(14.907,-15.000){2}{\rule{0.263pt}{0.400pt}}
\multiput(1701.00,446.92)(0.531,-0.494){27}{\rule{0.527pt}{0.119pt}}
\multiput(1701.00,447.17)(14.907,-15.000){2}{\rule{0.263pt}{0.400pt}}
\multiput(1717.00,431.92)(0.529,-0.494){29}{\rule{0.525pt}{0.119pt}}
\multiput(1717.00,432.17)(15.910,-16.000){2}{\rule{0.263pt}{0.400pt}}
\multiput(1734.58,414.82)(0.494,-0.529){29}{\rule{0.119pt}{0.525pt}}
\multiput(1733.17,415.91)(16.000,-15.910){2}{\rule{0.400pt}{0.263pt}}
\multiput(1750.58,397.82)(0.494,-0.529){29}{\rule{0.119pt}{0.525pt}}
\multiput(1749.17,398.91)(16.000,-15.910){2}{\rule{0.400pt}{0.263pt}}
\multiput(1766.58,380.82)(0.494,-0.529){29}{\rule{0.119pt}{0.525pt}}
\multiput(1765.17,381.91)(16.000,-15.910){2}{\rule{0.400pt}{0.263pt}}
\multiput(1782.58,363.72)(0.494,-0.561){29}{\rule{0.119pt}{0.550pt}}
\multiput(1781.17,364.86)(16.000,-16.858){2}{\rule{0.400pt}{0.275pt}}
\multiput(1798.58,345.72)(0.494,-0.561){29}{\rule{0.119pt}{0.550pt}}
\multiput(1797.17,346.86)(16.000,-16.858){2}{\rule{0.400pt}{0.275pt}}
\put(1074.0,669.0){\rule[-0.200pt]{3.854pt}{0.400pt}}
\end{picture}